\documentclass[aps,prb,preprint,onecolumn,amsmath,amssymb,superscriptaddress,eqsecnum,floatfix,scrartcl]{revtex4-1}
\pdfoutput=1      
\usepackage{amssymb}  
\usepackage{color,soul} 
\usepackage{graphicx}  
\usepackage[tight]{subfigure}  
\usepackage{mathrsfs}
\usepackage[pdftex,pdfusetitle,bookmarks=true,colorlinks=true,citecolor=blue,urlcolor=blue,linkcolor=magenta]{hyperref}
\usepackage{hypcap}

\DeclareMathOperator{\Ima}{Im}
\DeclareMathOperator{\Tr}{Tr}
\newcommand{\thg}[2]{\theta\!\left[{#1 \atop #2}\right]\hspace{-.3em}}

\begin{document}

\title{Two-cylinder entanglement entropy under a twist}    
\author{Xiao Chen}   
\email{xchen@kitp.ucsb.edu}
\affiliation{Kavli Institute for Theoretical Physics, University of California at Santa Barbara, CA 93106, USA}
\affiliation{Department of Physics and Institute for Condensed Matter Theory, University of Illinois at Urbana-Champaign, 1110 West Green Street, Urbana, IL 61801-3080, USA}

\author{William Witczak-Krempa} 
\email{w.witczak-krempa@umontreal.ca} 
\affiliation{D\'epartement de Physique, Universit\'e de Montr\'eal, Montr\'eal, Qu\'ebec, H3C 3J7, Canada}
\affiliation{Department of Physics, Harvard University, Cambridge MA 02138, USA}    

\author{Thomas Faulkner}
\email{tomf@illinois.edu}
\affiliation{Department of Physics and Institute for Condensed Matter Theory, University of Illinois at Urbana-Champaign, 1110 West Green Street, Urbana, IL 61801-3080, USA}
 
\author{Eduardo Fradkin}
\email{efradkin@illinois.edu}
\affiliation{Department of Physics and Institute for Condensed Matter Theory, University of Illinois at Urbana-Champaign, 1110 West Green Street, Urbana, IL 61801-3080, USA}

\date{\today} 

\begin{abstract}      
We study the von Neumann and R\'enyi entanglement entropy (EE) of scale-invariant theories defined on tori in 2+1 and 3+1 spacetime dimensions. We focus on spatial bi-partitions of the torus into two cylinders, and allow for twisted boundary conditions along the non-contractible cycles. Various analytical and numerical results are obtained for the universal EE of the relativistic boson and Dirac fermion conformal field theories (CFTs), and for the fermionic quadratic band touching and the boson with $z=2$ Lifshitz scaling. The shape dependence of the EE clearly distinguishes these theories, although intriguing similarities are found in certain limits. We also study the evolution 
of the EE when a mass is introduced to detune the system from its scale-invariant point, by employing a renormalized EE that goes beyond a naive subtraction of the area law. In certain cases we find 
non-monotonic behavior of the torus EE under RG flow, which distinguishes it from the EE of a disk. 
\end{abstract}

\maketitle   
\tableofcontents      

\section{Introduction}   
The entanglement entropy (EE) is a 
non-local quantity which, when properly treated, 
can capture the long-distance properties of correlated quantum many-body states  and quantum field theories. 
Given its non-local nature, the behavior of the EE is not as well understood as the correlators of local operators in quantum field theory 
and, in particular, how they are related to each other. Moreover, due to the non-local character of EE it may capture global properties 
which may not be accessible by  measurements of local operators.

The most complete and detailed understanding to date of the behavior of EE is in $1+1$-dimensional conformal field theories (CFT), 
which describe the universal behavior of systems at quantum criticality. In these CFTs the von Neumann EE 
of a single interval in the ground state has a universal logarithmic dependence on the length of the interval.   
The prefactor of the logarithm yields the central charge $c$ for the underlying CFT, 
which is the most important quantity to characterize the theory.\cite{callan1994, Holzhey-1994,calabrese_entanglement_2004,Vidal-2003}  
Furthermore, a related quantity, the mutual information of two disjoint intervals, depends not only on the central charge but also on the operator content 
of the CFT.\cite{Calabrese-2010} These results connecting the central charge of a  CFT to the scaling of the EE  suggested that the scaling of EE  
may be related to  definitions of a central charge in higher dimensions,\cite{Cardy-1988}  
and  to a generalization of  the Zamolodchikov  $c$-theorem\cite{Zamolodchikov-1986} for renormalization group flows.

Much less is known of the behavior of the EE in higher-dimensional field theories.   
In space dimensions $d>1$, the von Neumann EE of a local field theory on a finite but macroscopically large region of space  
satisfies the ``area law'' and 
scales with the size of the boundary of the observed region.\cite{Srednicki-1993,Bombelli-1986} 
In a local quantum field theory this behavior, reminiscent of the area law of the Bekenstein-Hawking black hole entropy,\cite{Bekenstein-1973,Hawking-1975}
is governed by a cut-off dependent non-universal prefactor, and
any universal (independent of the UV cutoff) behavior of the EE should be present in sub-leading corrections to the area law. 

Beyond the area law, many of the general results for the scaling of the EE in higher-dimensional CFTs are known primarily from the holographic EE of 
Ryu and Takayanagi.\cite{ryu2006a,ryu2006b} 
In higher-dimensional theories the geometry of the entangling regions (and the topology of space) plays a richer role. 
In 3+1 dimensions this richer structure allows for universal subleading logarithmic terms in the EE, even for
smooth surfaces.\cite{ryu2006b,Fursaev-2006,solodukhin2008entanglement} Such terms are present if the entangling regions are spheres,\cite{Casini2010,casini_towards_2011}  and have an universal coefficient related to the two central charges of the trace anomaly of the energy-momentum tensor.\cite{Cardy-1988} 

In this paper we will consider the finite universal terms in the scaling of the EE  both in 2+1 and 3+1 dimensions, 
where not many results are available. In particular, we will study the variation of this finite universal term under the twist boundary conditions.
From general arguments for local field theories, relativistic or not, we know that the leading term obeys the area law. 
In 2+1 dimensions, for an entangling region with a smooth boundary, 
one expects on dimensional grounds that the subleading term should be a finite  
scale-invariant function of the aspect ratio(s) of the entangling region. The first example of this behavior was found in topological phases of matter in 
2+1 dimensions, which are states with large-scale entanglement. In topological phases, and in topological quantum field theories such as Chern-Simons gauge theory,\cite{Witten-1989} the finite term of the von Neumann EE is a constant 
which is given in terms of  topological invariants of the underlying topological field theory
and on the topology of the observed region.\cite{kitaev_topological_2006,levin2006,Dong-2008}

For a 2+1-dimensional scale-invariant system, the EE of a region in the ground state satisfies the 
area law with a finite subleading correction, which, in general, depends on the shape of the region. 
This problem has been investigated in several field theories. 
One such theory is the quantum Lifshitz model,\cite{Ardonne-2004} which is a free compactified scalar field in 2+1 dimensions with dynamic critical 
exponent $z=2$ that describes the quantum (multi) critical point of generalized 
quantum dimer models in two dimensions.\cite{Rokhsar1988,Moessner-2001c,Fradkin-2004,Castelnovo-2005} 
In this model the von Neumann EE has a finite universal subleading term which depends on the compactification radius of the scalar field and of the aspect ratio 
(and the geometry) of the entangling surface. \cite{fradkin_entanglement_2006,hsu_universal_2009,stephan_shannon_2009,Hsu-2010,oshikawa_boundary_2010,stephan_renyi_2012,stephan_entanglement_2013,Zhou2016} A logarithmic dependence on the size of the region was found for entangling surfaces with corners (or cusps).\cite{fradkin_entanglement_2006,zaletel_logarithmic_2011} 

For relativistic CFTs in 2+1 dimensions, this scaling of the EE  has been shown to hold within the $\epsilon$-expansion for an O$(N)$ scalar field at the 
Wilson-Fisher fixed point for a partition of the $(3-\epsilon)$-dimensional space into two half-spaces with a planar entangling surface.\cite{ Metlitski2009} 
In the limit $N\to\infty$, the von Neumann EE was recently obtained for general entangling regions.\cite{Whitsitt-2016}
Moreover, for a circular entangling surface, the universal subleading term of the von Neumann EE for a disk, which in this context is usually called $-F$,  
is a universal constant which behaves much like   
the central charge in 1+1-dimensional CFT in that there is an appropriately defined function $\mathcal{F}(R)$ which is monotonically decreasing along a renormalization group flow from the UV to the IR and
agrees with the constant value of $F_{\rm UV,IR}$ at the respective fixed points. 
This result is known as the $F$-theorem.\cite{casini-huerta-2012,casini_mutual_2015}
In its strong form it was postulated within the AdS/CFT correspondence.\cite{Liu:2012eea}
The weaker version $F_{\rm UV} > F_{\rm IR}$ has been first proposed in a holographic context\cite{Myers-2010}
and subsequently verified to hold within supersymmetric 
models \cite{jafferis2011} and various other non-supersymmetric models.  
\cite{Klebanov-2011,Fei2015,Giombi-2015} 
In contrast to the case of smooth boundaries, for entangling regions with corners, a logarithmic dependence in the size of the region is  found also in explicit computations in relativistic models in 2+1 dimensions such as the the O$(2)$ and O$(3)$ scalars at their Wilson-Fisher fixed points,\cite{Stoudenmire2014,Kallin2014} the free massless scalar and the free massless Dirac fields,\cite{Casini_rev,Helmes16} and by the AdS/CFT correspondence.\cite{Hirata}     A general
argument was given for the coefficient of this log for opening angles close to $\pi$ in 
any relativistic CFT.\cite{bueno2015,bueno2015b,Faulkner2016}

\begin{figure}
\centering
\includegraphics[width=.62\textwidth]{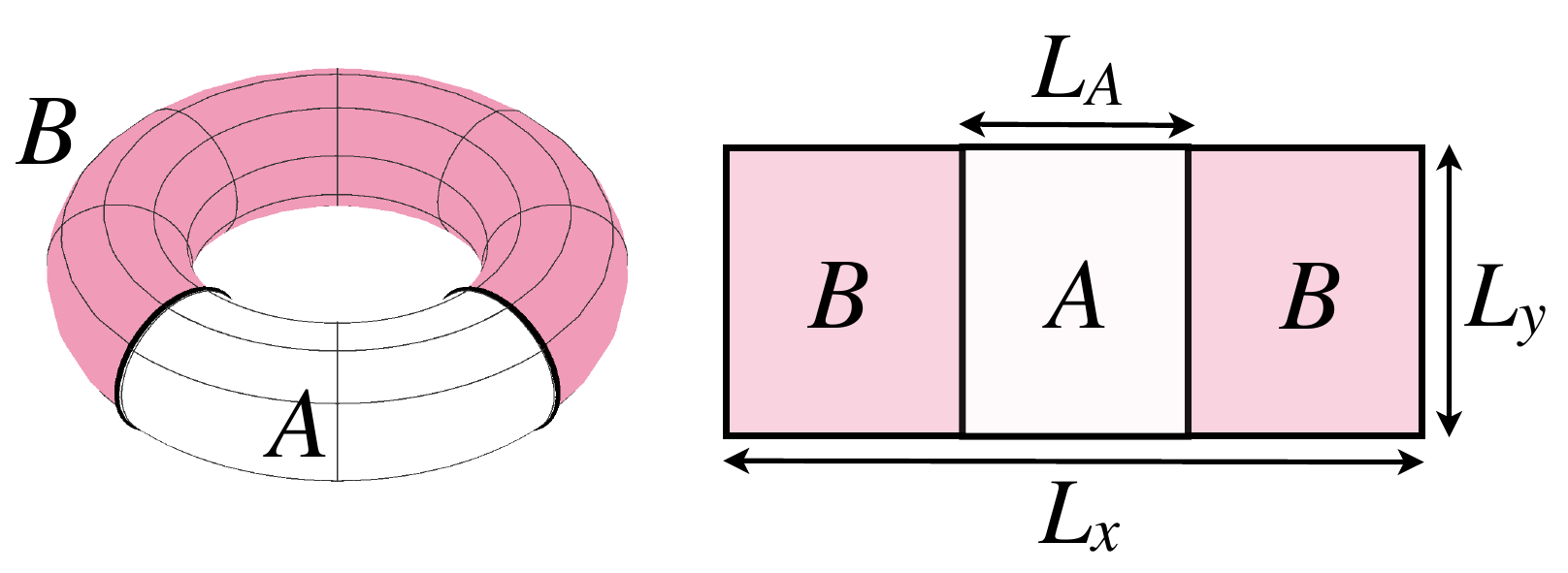}
\caption{The torus is divided into two cylinders $A$ and $B$ of size $L_A \times L_y$ and $(L_x-L_A)\times L_y$. The aspect ratios 
$u=L_A/L_x$ and $b=L_x/L_y$ characterize the geometry. }
\label{fig:schematic}
\end{figure}

Here we will consider the ground state of scale-invariant theories in 2+1 and 3+1 dimensions on a torus and investigate the scaling of entanglement when the torus is 
partitioned into two cylinders. In 2+1 dimensions, the torus has circumferences $L_x$ and $L_y$ and the observed cylindrical region $A$ has length $L_A$, as shown in Fig.\ref{fig:schematic}. 
The R\'enyi EE for the cylindrical region $A$, of aspect ration $u=L_A/L_x$ ,
satisfies the area law with a finite subleading correction, \cite{hsu_universal_2009,stephan_entanglement_2013,Inglis2013,chen_scaling_2014,krempa2016,Chojnacki2016}
\begin{align}
  S_n=\alpha \frac{2L_y}{\epsilon}-J_n(u;b) +\dotsb
\end{align} 
where $n$ is the R\'enyi index, and $b=L_x/L_y$.  
The leading area law term arises from short-range entanglement localized to the boundary between $A$ and $B$, 
and is proportional to the length of the boundary, $2L_y$. The coefficient $\alpha$ is non-universal, being cutoff dependent. 
We will discuss the behavior of the finite universal term $J_n(u;b)$ in several theories on a torus with \emph{twisted} boundary conditions, 
which is allowed by the non-trivial topology.   

We emphasize that $J_n(u;b)$ is a \emph{universal} scaling function that depends on the two 
aspect ratios: $b=L_x/L_y$ and $u=L_A/L_x$, and varies from theory to theory. It thus acts as a non-trivial fingerprint for quantum systems.  
It has been studied in several model systems and there are at least three
analytical expressions for it. One expression was derived in the quantum Lifshitz model describing a compact free boson with 
dynamics exponent $z=2$.\cite{Hsu-2010,stephan_renyi_2012,stephan_entanglement_2013} The ground state of this model has the conformal invariance 
in spatial direction and therefore $J_n(u,b)$ can be analytically constructed in terms of partition functions for two-dimensional CFTs. An explicit expression on the torus has been given for $n\geq 2$.\cite{stephan_entanglement_2013} The other expression was derived holographically using the Ryu-Takayanagi formula, valid within the AdS/CFT correspondence,  and is expected to yield the EE for certain strongly interacting CFTs.\cite{ryu2006a, ryu2006b,chen_scaling_2014} 
The third instance,\cite{krempa2016} which is the simplest, is obtained within a toy theory called the Extensive Mutual Information model.\cite{Casini08,Swingle10} 
Although these three expressions are different, they share similar properties and take the same scaling form in various limits. 
The $J$-function can also be computed numerically for 
some simple theories.\cite{chen_scaling_2014,krempa2016,Chojnacki2016,Whitsitt-2016} The $n=2$ scaling function has also been determined by quantum Monte Carlo simulations 
at the quantum critical point of the two-dimensional Ising model in a transverse field by Inglis and Melko.\cite{Inglis2013}      
Although the quantum critical behavior of this model is described by the relativistic real scalar field at the Wilson-Fisher fixed point, Inglis and Melko found that the finite term of the second R\'enyi entropy is (surprisingly) well approximated by the scaling function of the quantum Lifshitz model, derived by Stephan and coworkers.\cite{stephan_entanglement_2013}      

In this paper we study both analytically and numerically the behavior of the scaling function $J(u;b)$ for two massless relativistic models in 2+1 dimensions, the relativistic massless scalar field (which we refer to as the ``free boson model'') and the massless Dirac field. We also study a free massless boson model with dynamical exponent $z=2$ and a free fermion  massless model with dynamical exponent $z=2$ (the quadratic band touching model\cite{sun-2009}), both also in 2+1 dimensions.
Although all four models are massless, and hence define scale invariant systems in 2+1 dimensions, they describe very different types of fixed points. Indeed, in 2+1 dimensions the free relativistic massless 
scalar field is unstable in the IR,    
and under a $\lambda \phi^4$ perturbation  flows to the Wilson-Fisher fixed point. 
Similarly, the free massless Dirac field is an IR stable fixed point and, as such, it defines a stable phase of matter. In contrast, both $z=2$ theories, the free boson and the quadratic band touching fermion model describe systems at marginality (are ``asymptotically free'', and hence perturbatively renormalizable). These differences make the comparisons of the scaling functions interesting. In particular, here we focus on the toroidal geometry with twisted boundary conditions, and investigate the IR flows of the EE in these geometries and compare its behavior in
the free (non-compact) complex scalar field and the free (also non-compact) Dirac fermion.  

We will not discuss here the interesting case of the compact relativistic  boson (which should be regarded as a Goldstone boson of a spontaneously broken $U(1)$ symmetry and, hence, describes a theory at the IR stable fixed point).
The entanglement properties of this model have been discussed in the literature analytically only for the geometries of disks \cite{agon2014disk,metlitski2011entanglement} and numerically on cylinders.\cite{Kallin-2011} Likewise, we will not discuss the von Neumann EE for the compactified quantum Lifshitz model 
on a torus. While its entanglement properties have  been extensively discussed in several geometries,  for the torus the only analytical results 
available are for R\'enyi entropies with $n\geq 2$ and not the von Neumann case, $n=1$.\cite{stephan_entanglement_2013}

\subsection{Summary of the main results}  

 In this paper, we  study the function $J_n(u;b)$ in several free field theories. For the free relativistic boson CFT and the Dirac fermion CFT in $2+1$ dimensions, although the complete analytical form for $J_n(u;b)$ is unknown, at around $u=1/2$, $J_n$ can still be analytically obtained in the thin torus limit.\cite{Metlitski2009} As $b$ increases, $J_n(1/2)$ converges to a constant which only depends on the boundary condition in $y$ direction. Here $J_n(1/2)$ is the shorthand notation for $J_n(1/2; b\to\infty)$. We also extend this calculation to $3+1$ dimensions and calculate $J_n(1/2)$ in the thin torus limit for various boundary conditions. We verify our analytical expression by performing numerical calculation on lattice models.

We further explore the monotonicity of $J_n(1/2)$ and the connection with the $F$-theorem, which states that for the $2+1$ dimensional relativistic CFT, the subleading correction term of the von Neumann EE for a disk is a universal constant and can serve as an RG monotone.\cite{jafferis2011,casini_towards_2011, casini_mutual_2015,Giombi-2015} We explicitly add a mass term in these free systems and define some renormalized EE so that it is equal to $J(1/2)$ when mass is zero and approaches to zero as mass increases. However, no matter how we define renormalized EE, it cannot be an RG monotone. This is because as we change the boundary condition, $J_n(1/2)$ can take both positive and negative values. A similar issue happens in $3+1$ dimensions, where $J_n(1/2)$ changes sign as we change the boundary condition.  

We also study the  $z=2$ free boson model in $2+1$ dimensions with twisted boundary conditions. Since the ground state wave function has conformal invariance, we can obtain the complete $J(u,b)$ by applying the replica trick method directly on the ground state wave function. In the wide torus limit $b\to 0$, $J_n(u)$ can be exactly mapped to the corner correction $J(\theta)$ defined on an infinite plane. In the thin torus limit, $J_n(1/2)$ takes the same form as that for free boson CFT and Dirac fermion CFT up to a prefactor.
For completeness, we also study the fermionic quadratic band touching model,\cite{sun-2009} a critical fermionic system with $z=2$, and analyze $J_n(u)$ function numerically in various limits. The main difference from the other three models in $2+1$d is that $J_n(1/2)$ always equal to zero for any $b$.

For these $2+1$ dimensional non-interacting scale-invariant models, 
$J_n(u)$ in the thin torus limit $L_y\to 0$ reads 
\begin{align} \label{gamma-all}
  2\gamma_n =
  \begin{cases}
    \frac{1}{3} (1+\frac{1}{n})\log(2\sin(\pi\lambda)), & \mbox{free complex boson CFT}\\
    \frac{1}{6}(1+\frac{1}{n})\log(2\sin(\pi\lambda)), & \mbox{Dirac fermion CFT}\\
    2\log(2\sin(\pi\lambda)), & \mbox{$z=2$ complex boson}\\
    0, & \mbox{fermionic quadratic band touching}
  \end{cases}
\end{align}
where $\lambda\in (0, 1)$ denotes the twist along the $y$ direction\ and in the bosonic models, it is defined as follows
\begin{align}
  \phi(x,y+L_y)=e^{i 2\pi \lambda}\, \phi(x,y)
\end{align}
In the fermionic models, $\lambda$ can be defined in the same way.  $\lambda=1/2$ corresponds to anti-periodicity.
Here, $\gamma_n$ is (minus) the universal EE of a semi-infinite cylindrical region obtained by
bi-partioning an infinite cylinder. 
For the first three theories, $\gamma_n$ has the  
same dependence on the twist $\lambda$ up to a prefactor and will diverge in the limit $\lambda\to 0$. Interestingly, the \emph{complex} boson
value of $\gamma_n$ is twice that of the Dirac fermion, and the quadratic band touching has a vanishing
universal contribution in this limit. We note in passing that in all four theories, $\gamma_n$ vanishes identically for the special twist parameter $\lambda=1/6$. 

For the $3+1$ dimensional complex scalar CFT, in the thin torus limit with $L_y, L_z\to 0$, 
\begin{align}
2\gamma^{3d}_n=\frac{1}{6}\left(1+\frac{1}{n}\right) \log\left( \frac{ \thg{\lambda_2-\frac{1}{2}}{\lambda_1-\frac{1}{2}}(\tau)}{\eta(\tau)}
\frac{ \thg{\lambda_2-\frac{1}{2}}{-\lambda_1+\frac{1}{2}}(\tau) }{\eta(\tau)}\right) 
\end{align}
where $\tau=ir=i L_y/L_z$ is the modular parameter of each of the two boundaries of $A$, which are 2-tori.
$\theta{\alpha\brack\beta}(\tau)$ is given in terms of a Jacobi theta function, Eq.~\eqref{elliptic}.    

The structure of this paper is as follows. We first discuss the two cylinder EE for 
relativistic free field theory with twisted boundary condition in $2+1$ dimensions  
in Sec.~\ref{2d_z_1}. We focus on the thin torus limit and define the renormalized EE.  
Then we compute two-cylinder EE defined on cylinder and torus for $z=2$ free boson theory with twisted boundary condition in Sec.~\ref{2d_z_2}. We consider various limits in both cases and for comparison, we also numerically study the fermionic quadratic band touching model.   We further study the two-cylinder EE for relativistic free field theory with twisted boundary condition in $3+1$ dimensions in Sec.~\ref{3d_z_1}. We summarize and conclude in Sec.~\ref{conclusion}. The appendices are devoted to details of the calculations and techniques used in this paper. 

\section{General properties of torus entanglement in two dimensions}

We review the basic properties of the universal torus function $J_n(u;b)$. Our present discussion is
concerned with the thermodynamic limit of $J_n$, where lattice effects have been extrapolated away. 
First, the R\'enyi entanglement entropies of the cylindrical regions $A$ and $B$ are equal, since we work with the ground state, which is pure.
This leads to the reflection symmetry: $J_n(1-u)=J_n(u)$.
Further, the strong subadditive property of the EE implies\cite{krempa2016} that $J_1(u)$ is a decreasing convex function of 
$u$ on the interval $(0,1/2]$, for any value of $b$ and any  
choice of boundary conditions. 
For a fixed aspect ratio $b$, $J_1(1/2)$ is thus the smallest value of $J_1(u)$.  
These properties can be clearly seen in Figs.~\ref{boson_torus} and \ref{Dirac_torus}. 
We now consider various limits where we can make exact statements.
\begin{figure}
  \centering
  \includegraphics[width=.5\textwidth]{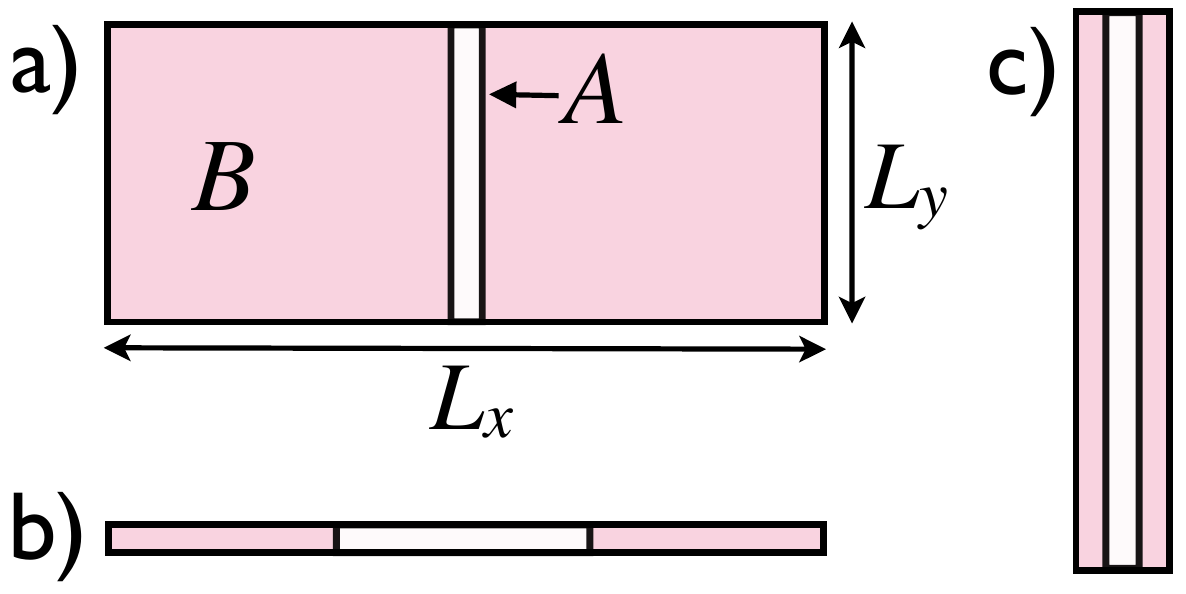}
  \caption{Sketch of various limits of the entangling geometry on the two-dimensional torus: 
a) thin slice ($L_A\to 0$), b) thin torus ($L_y\to 0$)
and c) wide torus ($L_y\to \infty$). In taking these limits, the other lengths are fixed.
This means that a) $u\to 0$ and $b$ finite, b) $u$ finite and $b\to\infty$, c) $u$ finite and $b\to 0$.}   
  \label{fig:shapes}
\end{figure}

\subsubsection{Thin slice limit}

An important limit is the so-called thin slice limit $u\to 0$, Fig.~\ref{fig:shapes} (a), where the universal term diverges as
\begin{align}
  J_n(u\to 0;b) = \frac{\kappa_n}{u}\,,
\end{align}
where $\kappa_n$ is a universal coefficient characterizing the theory. Interestingly, this is the same coefficient that dictates
the universal R\'enyi entropy of long strip living in the infinite plane: 
$S_{\rm strip}=BL/\delta - \kappa L/w$, where $w$ is the strip's width and $L$ is
a scale used to regulate the large-distance divergence. Alternatively, one can consider the EE per unit length $S_{\rm strip}/L$. 
The reason for the appearance of the same $\kappa$ in both geometries is that the boundary conditions along the $x$ and $y$ 
cycles of the torus will not influence the universal EE as $u\to 0$. 
(We assume, as is generically the case, that there are no zero modes in the compactified geometry.)    

\subsubsection{Thin torus limit}

As the name suggests, we take $L_y\to 0$, while keeping $L_A,L_x$ fixed, Fig.\ref{fig:shapes} (b) . This implies that the aspect ratio diverges $b\to\infty$, 
while $0<u<1$ remains fixed. In this case, the universal
R\'enyi entropy will tend to a constant,
\begin{align}
  J_n(u;b\to \infty) = 2\gamma_n\,,
\end{align}
where $\gamma_n$ is independent of all length scales. This saturation comes about because
we consider generic theories/boundary conditions precluding zero modes, so that the theory on the torus possesses a large gap $\sim 1/L_y$,
and thus becomes insensitive to the length scales $L_x,L_A\gg L_y$. Further, the universal part of the EE cannot depend 
on $L_y$ because no other length scale remains to form a dimensionless ratio (we work with the groundstate of a scale invariant system). 
In contrast, when zero modes are present (non generic), $\gamma_n$ will depend on the scales $L_x,L_A$ and the short-distance cutoff $\epsilon$,
as we illustrate in Sections \ref{2d_z_1} and \ref{3d_z_1} with the free boson and fermion with periodic boundary conditions.

\subsubsection{Wide torus limit} 

\begin{figure}[hbt]
\centering
\includegraphics[width=.7\textwidth]{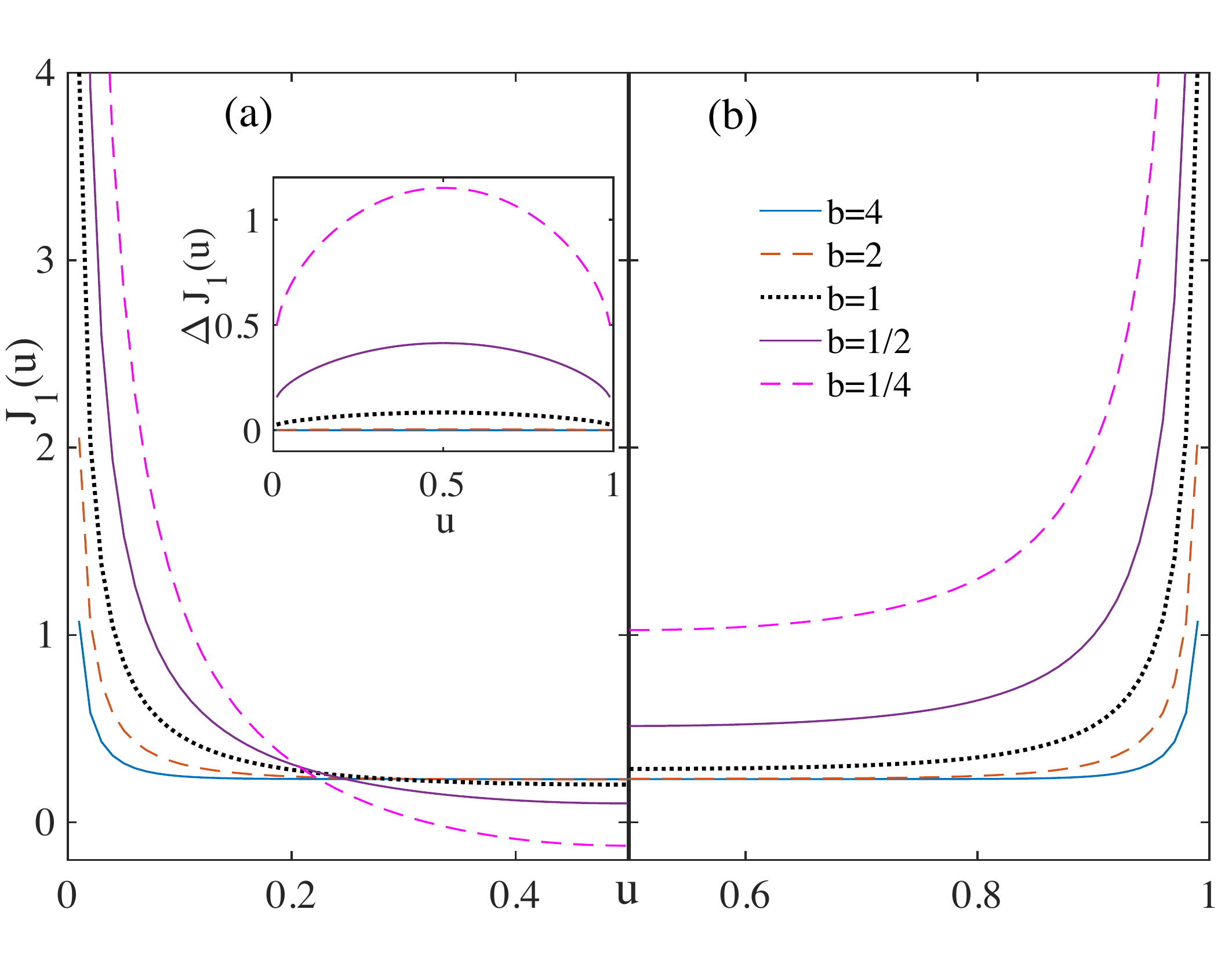}
\caption{(Color online) (a) The universal EE on the torus $J_1(u)$ for the non-interacting real boson 
CFT for various $b=L_x/L_y$ with anti-periodic boundary conditions (BCs) in the $y$ direction 
and periodic BCs in the $x$ direction: $(\lambda_x,\lambda)=(0,1/2)$. 
The result is doubled for a complex boson. We use $L_x=400$ lattice sites in the $x$-direction.  
(b) $J_1(u)$ with anti-periodic BCs in both directions: $\lambda_x=\lambda=1/2$. The other parameters are the same as in (a). Since $J_1(u)$ is symmetric around $u=1/2$, we only show $u\in (0, 1/2]$ in (a) and $u\in [1/2, 1)$ in (b). 
The inset shows the difference, $\Delta J_1(u)$, between the two choices of BCs.} 
\label{boson_torus} 
\end{figure}

We take the opposite limit, $b\to 0$, by sending $L_y$ to infinity, but again keeping $L_A,L_x$ fixed, Fig.\ref{fig:shapes} (c).
In this case, the EE will be dominated by the diverging length scale $L_y$. By translation invariance along the
$y$-direction, $J_n$ is expected to scale extensively with $L_y$:
\begin{align}
  J_n(u;b\to 0) = L_y \cdot \frac{f_n(u)}{L_x} = \frac1 b \cdot f_n(u)
\end{align}
where we have made $f_n(u)$, which is independent of $b$, dimensionless by factorizing $1/L_x$. The growth of $|J_n|$ with decreasing $b$
can be observed in Figs.~\ref{boson_torus} and \ref{fig:cons_term} for the free boson and Dirac fermion CFTs, respectively. 
Further, in a holographic calculation\cite{chen_scaling_2014} for an interacting CFT,  
it was found that the relation $J_1=f_1(u)/b$ is exactly obeyed \emph{for all} $b<1$.    

\begin{figure}[hbt]
\centering
\vspace*{0cm}
\includegraphics[scale=.5]{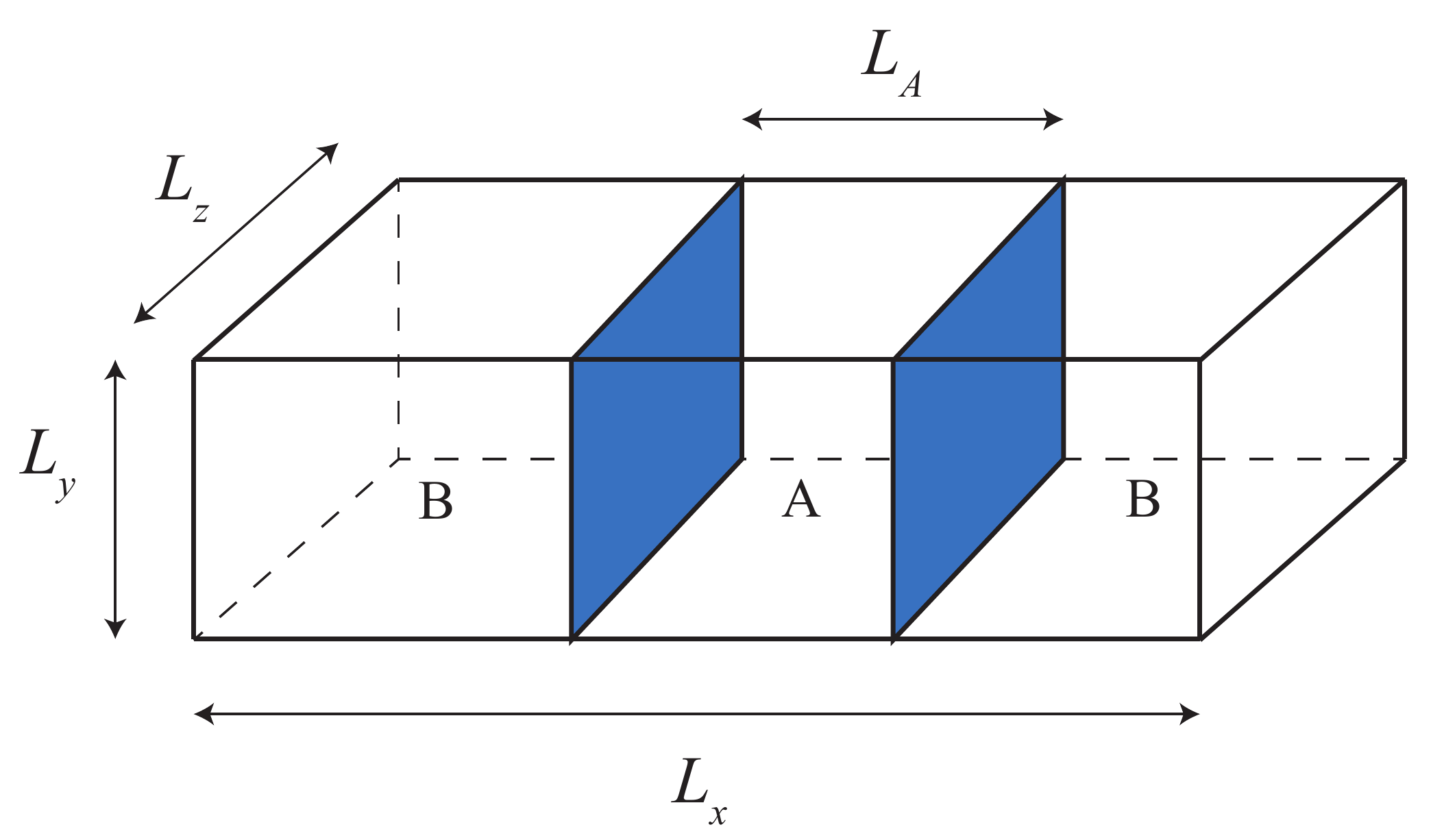}
\caption{Three dimensional torus, with opposite faces of the box being identified. Region $A$
is a cylinder of length $L_A$, with 2 boundaries (blue/shaded).} 
\label{3d_torus}
\end{figure}

The $J$-function can be generalized to $3+1$ dimensional theories defined on a three-torus, as shown in Fig.\ref{3d_torus}.\cite{krempa2016} 
In this case, $J_n$ depends on the aspect ratio of the subsystem $A$, $u=L_A/L_x$, as well as on the two aspect ratios of the
torus: $b_1=L_x/L_y$ and $b_2=L_x/L_z$. In the thin slice limit $u\to 0$, $J_n(u)\to 1/(u^2b_1b_2)$,\cite{krempa2016} 
while in the thin torus limit $b_1,b_2\to\infty$, $J_n$ will saturate to $\gamma_n^{\rm 3d}(L_y/L_z)$, which only depends on the aspect ratio of
the boundary of region $A$. 

\section{Two-cylinder entropy for relativistic free field theory}
\label{2d_z_1} 
The continuum Hamiltonian for the relativistic boson in $2+1$d is
\begin{align}
  H=\frac{1}{2}\int dxdy \left(\Pi(x,y)^2+(\nabla\phi(x,y))^2+m^2\phi(x,y)^2\right),
  \label{H_boson_z1}
\end{align} 
where the integral is over the $L_x\times L_y$ torus. 
Since the $y$ direction is compactified into a circle of length $L_y$, we can decompose $\phi(x,y)$ 
into discrete Fourier modes in the $y$ direction 
\begin{align}
\phi(x,y)=\sum_{k_y} \frac{e^{-ik_y y}}{\sqrt{L_y}}\phi_{k_y}(x)
\end{align}
with quantized momentum $k_y$:
\begin{align}
  k_y=\frac{2\pi(p+\lambda)}{L_y}\,,
\end{align}
where $p\in \mathbb{Z}$ and $\lambda$ parametrizes the twist along the $y$ direction. In other words, we have 
\begin{align}
  \phi(x,y+L_y)=e^{i 2\pi \lambda}\, \phi(x,y)
\label{phi_twist}
\end{align}
The twist in the $x$-direction is $\lambda_x$.
For a real scalar field, $\phi(x,y)$ is a  Hermitian operator, so $\lambda,\lambda_x$ 
can only take the values $0$ or $1/2$, which correspond to periodic and  
anti-periodic boundary conditions, respectively. 
For a real bosonic field, the Fourier modes need to satisfy $\phi^{\dag}_{k_y}(x)=\phi_{-k_y}(x)$, therefore Eq.\eqref{H_boson_z1} can be rewritten as 
\begin{align}
H=\sum_{k_y=-\infty}^{\infty}\frac{1}{2}\int dx \left[\Pi_{k_y}^{\dag}(x)\Pi_{k_y}(x)+(\partial_x\phi^{\dag}_{k_y}(x))(\partial_x\phi_{k_y}(x))+(m^2+k_y^2)\phi^{\dag}_{k_y}(x)\phi_{k_y}(x)\right]
\end{align}
Notice that only the $k_y\geq 0$ modes are independent degrees of freedom.

$\phi_{k_y}(x)$ can be further decomposed as follows:
\begin{align}
\phi_{k_y}(x)=\frac{1}{\sqrt{2}}(\phi_{k_y,1}+i\phi_{k_y,2})
\end{align}
The Hamiltonian becomes 
\begin{align}
H=&\sum_{k_y=-\infty, I=1,2}^{\infty}\frac{1}{4}\int dx\left[ \Pi^2_{k_y,I}(x)+(\partial_x\phi_{k_y,I})^2+(m^2+k_y^2)(\phi_{k_y,I})^2 \right]\nonumber\\
=&\sum_{k_y\geq 0}\frac{1}{2}\int dx\left[ \Pi^2_{k_y,1}(x)+(\partial_x\phi_{k_y,1})^2+(m^2+k_y^2)(\phi_{k_y,1})^2 \right]\nonumber\\
&+\sum_{k_y>0}\frac{1}{2}\int dx\left[ \Pi^2_{k_y,2}(x)+(\partial_x\phi_{k_y,2})^2+(m^2+k_y^2)(\phi_{k_y,2})^2 \right]
\label{H_boson}
\end{align}
where we use $\phi_{k_y,I}=\phi_{-k_y,I}$ ($I=1,2$), and the component $\phi_{0,2}=0$. Note that the $k_y=0$ mode exists only when $\lambda=0$. 
This Hamiltonian consists of a sum of decoupled $1+1$d free bosons, with effective mass $\sqrt{m^2+k_y^2}$. 

Similarly, for Dirac fermions in $2+1$d, the Hamiltonian on the torus can be expressed as a sum over an infinite number of $1+1$-dimensional 
massive Dirac fermions with a mass given by $\sqrt{m^2+k_y^2}$.

\begin{figure}[hbt]
\centering
\includegraphics[width=.7\textwidth]{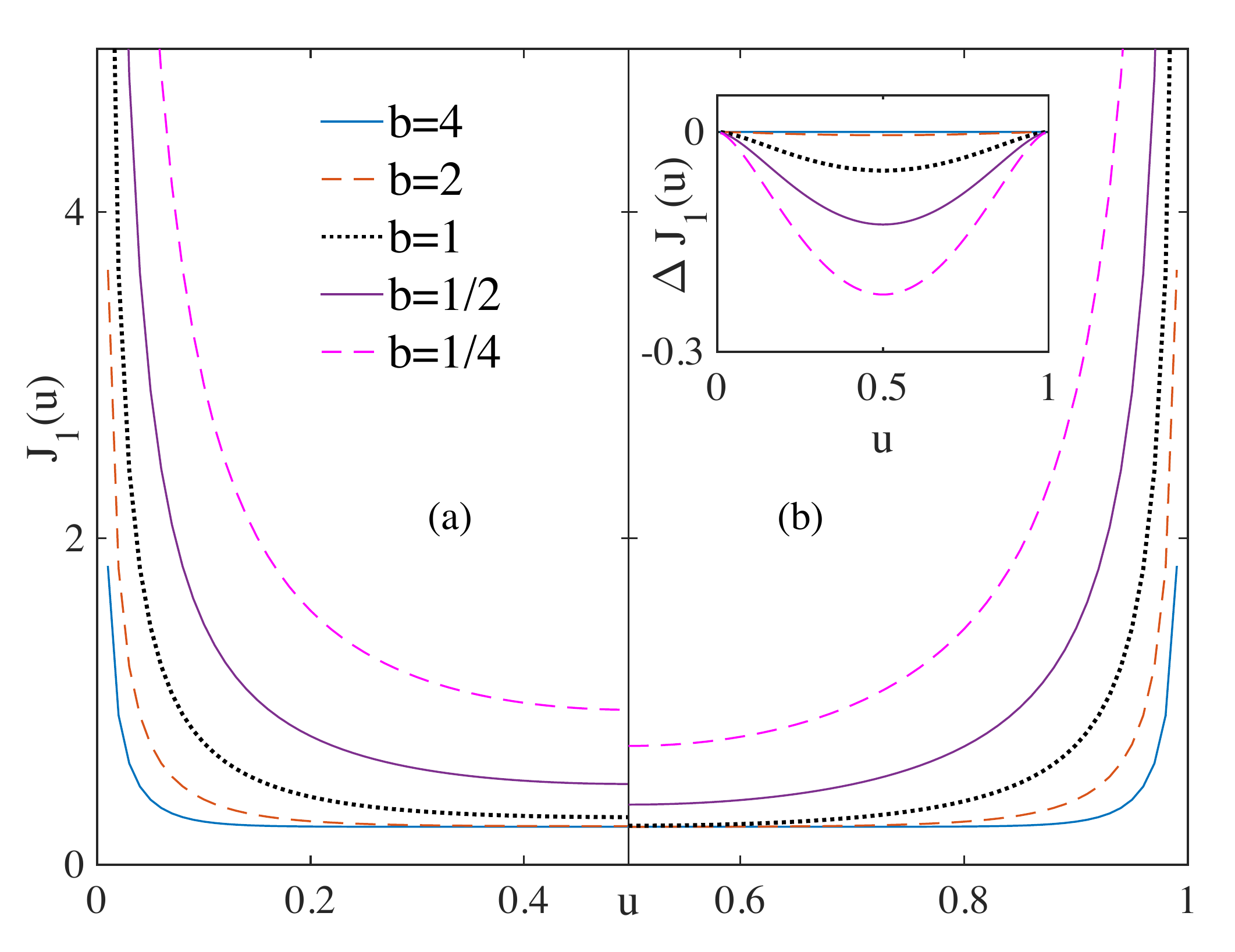}
\caption{(Color online) The subleading correction $J_1(u)$ of Dirac fermion for various $b=L_x/L_y$ with anti-periodic boundary condition 
in the $y$ direction and periodic boundary condition in the $x$ direction, $(\lambda_x,\lambda)=(0,1/2)$. $L_x=400$. (b) $J_1(u)$ with anti-periodic boundary conditions 
in both directions: $\lambda_x=\lambda=1/2$. The other parameters are the same as (a). The inset shows their difference $\Delta J_1(u)$ at various $b$.} 
\label{Dirac_torus}
\end{figure}

\subsection{The shape dependence}  
We bi-partition the torus into two cylinders as shown in Fig.~\ref{fig:schematic} and calculate the 
entanglement entropy between two cylinders.  Since a full analytical treatment of the shape dependence for the torus EE of relativistic bosons
and Dirac fermion is difficult, we compute the EE numerically by using the method explained in Appendix \ref{numerical_method}. By exploiting the decomposition into decoupled (1+1)-dimensional
massive chains described above, we are able to work with very large lattices. A similar analysis was previously done for the free Dirac fermion and free boson CFT, 
but only the quantity $-S(u)+S(1/2)= J(u)-J(1/2)$ was obtained, and only for a limited set of boundary conditions.\cite{chen_scaling_2014,krempa2016,Chojnacki2016}
Here, we determine the full $J(u;b)$, including the $J(1/2;b)$ part, by subtracting the area law contribution from $S$.  
The result is shown in Figs.~\ref{boson_torus} and \ref{Dirac_torus} for the boson and fermion, respectively.  
In those figures, two sets of boundary conditions are used: $(\lambda_x,\lambda)=(0,1/2)$ and $(1/2,1/2)$,
where $0$ means periodic while $1/2$ is anti-periodic.
As mentioned in the introduction, $J_1(u)$ is a decreasing convex function of $u$ on the interval $(0,1/2]$, for any value of $b$ and any
choice of boundary conditions. This follows from the strong subadditivity property of the EE. 
For a fixed aspect ratio $b$, the minimum of value of $J_1(u)$ is thus $J_1(1/2)$. $J_n(1/2;b)$ is plotted as function of $b$ in Fig.~\ref{fig:cons_term}
for both the fermion and boson, and R\'enyi indices $n=1,2$. 

An important limit is the so-called thin slice limit $u\to 0$, where the universal term diverges as
\begin{align}
  J_n(u\to 0;b) = \frac{\kappa_n}{u}
\end{align}
where $\kappa_n$ is a universal coefficient characterizing the theory. Interestingly, this is the same coefficient that dictates
the universal R\'enyi entropy of the long thin strip. The point is that the boundary conditions along the $x$ and $y$
directions will not alter the value of $\kappa$ (we assume that both boundary conditions are not simultaneously periodic
to avoid the zero mode).   

\subsection{Thin torus limit \& semi-infinite cylinder}

We examine another important limit, the thin torus, obtained by sending $b\to\infty$, while the ratio $u$ remains fixed.
In this case, the universal R\'enyi entropy saturates to a pure constant when $0<\lambda<1$, 
\begin{align}
  J_n(u;b\to\infty)\to 2\gamma_n(\lambda)\,, 
\end{align}
where $\gamma_n$ is the subleading term in the EE for the bi-partition of an infinite cylinder into
semi-infinite cylinders, as explained in the introduction. We note that the universal constant $\gamma_n$ only depends on the twist 
along the $y$-direction, $\lambda$. This becomes manifest in Figs.~\ref{boson_torus}-\ref{Dirac_torus}, where the left panels have $\lambda_x=0$,
while the right ones have $\lambda_x=1/2$. 
Their differences, $\Delta J_1(u)$, for the two choices of $x$-boundary conditions are shown in the insets. 
Indeed, at large $b$, $J_1(u)$ becomes insensitive to $\lambda_x$, and $\Delta J_1(u)\approx 0$. 

We can analytically compute $\gamma_n$ for all $n$ using the 1d decomposition given above, Eq.\eqref{H_boson}.
For a finite interval of length $L_A$, each $1+1$d massive chain, with effective mass $\sqrt{m^2+k_y^2}$, contributes an EE \cite{calabrese_entanglement_2004}
\begin{align}
  S_n^{\rm 1d}(k_y)=- \frac{1}{12} \left(1+\frac{1}{n}\right) \log\left[(m^2+k_y^2)\epsilon^2\right]
\end{align}
where the length of the interval is taken to be much larger than the inverse mass, $L_A\gg 1/\sqrt{m^2 +k_y^2}$; $\epsilon$ 
is the UV cutoff, and $n$ the R\'enyi index. The total EE is then   
\begin{align}
  S_n=-\sum_{k_y} \frac{1}{12} \left(1+\frac{1}{n}\right) \log\left[(m^2+k_y^2)\epsilon^2 \right]
\label{sum_1d}
\end{align}

\begin{figure}[hbt]
\centering
\includegraphics[width=.7\textwidth]{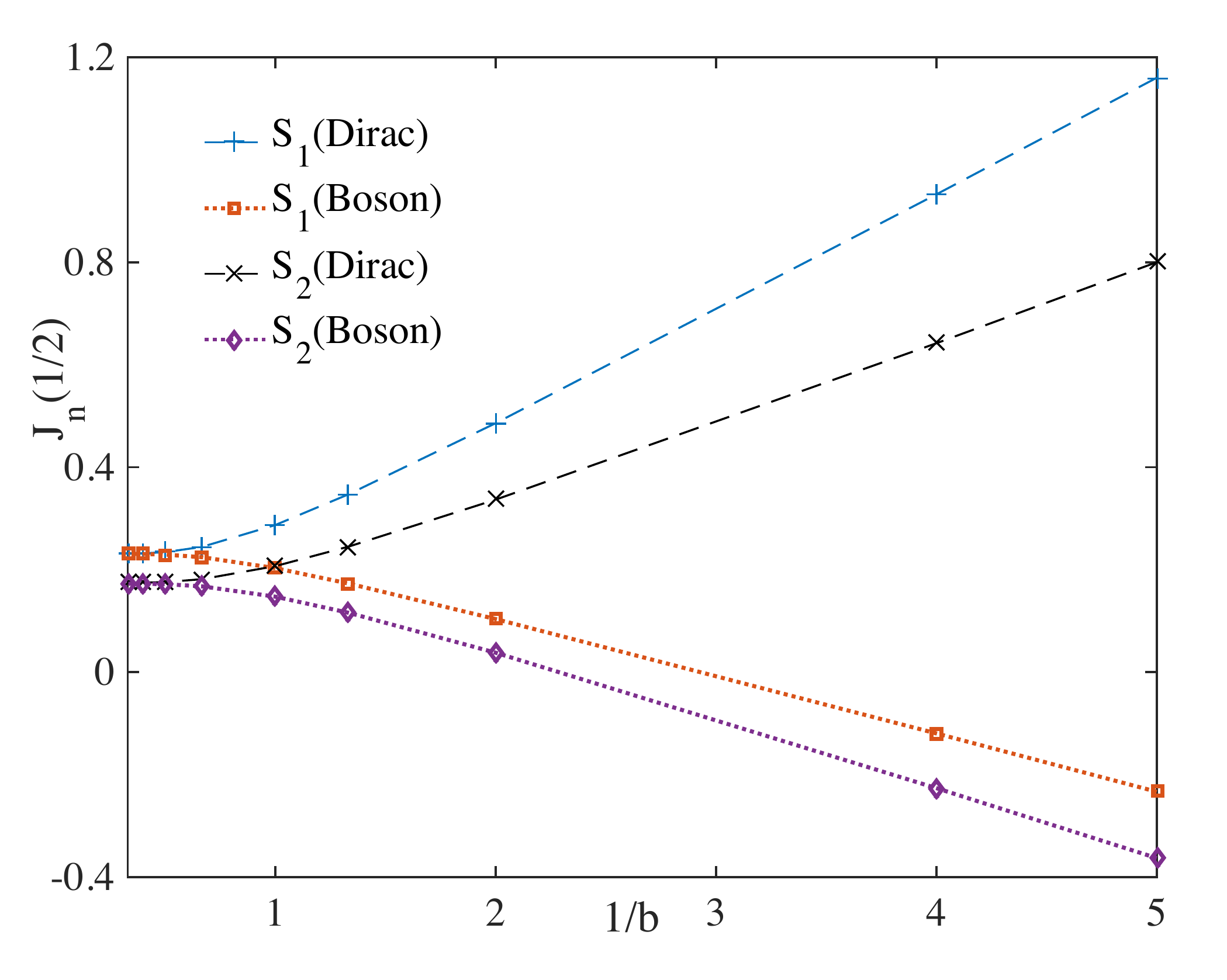}
\caption{(Color online) Numerical lattice calculation of $J_n(1/2;b)$ as a function of $1/b$ for a real boson and a 2-component Dirac fermion. 
We choose anti-periodic boundary conditions along $y$,  
and periodic boundary condition along $x$. When $n=1$, $J_1(1/2;b\gg 1)$ converges to 0.2310 
and when $n=2$, $J_2(1/2;b\gg 1)$ converges to 0.174. 
Every data point is obtained by increasing $L_y$ from $400$ to $720$ and subtracting the leading area law term.} 
\label{fig:cons_term}
\end{figure}

We first consider the massless case, $m=0$,
\begin{align}
  S_n=-\sum_{p\in\mathbb{Z}} \frac{1}{6} \left(1+\frac{1}{n}\right)\left(\log|2\pi (p+\lambda)|+\log\frac{\epsilon}{L_y}\right)
  \label{S_n_2d}
\end{align}
where $k_y=2\pi(p+\lambda)/L_y$. 
$\lambda$ denotes the twist boundary condition in the $y$ direction, $\phi(x, y+L_y)=e^{2\pi i\lambda}\phi(x, y)$. 
This expression can be regularized and we get
\begin{align}
  S_n= \alpha \frac{L_y}{\epsilon} - 2\gamma_n
\end{align} 
By using the Hurwitz zeta function regularization method discussed in Appendix \ref{hurwitz}, we have for $0<\lambda<1$
\begin{align}
2\gamma_n=\sum_{p\in\mathbb{Z}}\frac{1}{6} \left(1+\frac{1}{n}\right)\log|p+\lambda|=\frac{1}{6}\left(1+\frac{1}{n}\right) \log(2\sin(\pi\lambda)) 
\label{gamma_2d_z_1}
\end{align}
For the real boson field, $\lambda$ can only be $0$ or $1/2$. 
When $\lambda=1/2$, $2\gamma_1=\frac{\log 2}{3}=0.231$. This matches the numerical results shown in Fig.~\ref{fig:cons_term}. 
For the Dirac fermion, $\lambda$ can take values between $0$ and $1$, which corresponds to arbitrary 
twisted boundary condition and can be realized by coupling Dirac fermion to a $U(1)$ gauge field. 
In this case, $\gamma_n$ takes the same formula shown in Eq.\eqref{gamma_2d_z_1}. For a \emph{complex} boson, $\lambda\in (0,1)$ and $\gamma_n$ is twice of that for real boson field or Dirac fermion field. $\gamma_n$ in the thin torus 
limit takes the same form for both $z=1$ free boson and Dirac fermion model.

Strictly speaking, Eq.\eqref{gamma_2d_z_1} only works for $0<\lambda<1$. 
As $\lambda\to 0$, the effective mass of the $k_y=0$ $(1+1)$d mode vanishes and the use of Eq.\eqref{gamma_2d_z_1}
will yield an incorrect result. This issue can be resolved by separately treating the massive modes $k_y\neq 0$ and the zero mode with $k_y=0$.  
The $k_y=0$ zero mode contributes a subleading term to the EE that will depend on the aspect ratio
$u=L_A/L_x$, as well as on $L_x/\epsilon$, where $\epsilon$ is the UV cutoff. In addition, this subleading 
term \emph{will depend on the twist along the x-direction, $\lambda_x$.} 
Since the zero mode is not suppressed by a $\sim 1/L_y$ gap, it is sensitive to the entire geometry, including 
the boundary conditions along the ``long direction'', $x$. 
For the special case of periodic boundary conditions along $x$, $\lambda_x=0$, we obtain 
\begin{align} \label{eq:gam-zero}
2\gamma_n= -\frac{1}{6}\left(1+\frac{1}{n}\right) 
\log \left(\frac{L_x}{\pi\epsilon} \sin\left(\frac{\pi L_A}{L_x}\right)\right)   
\end{align} 
The massive modes, after regularization, contribute a finite non-universal constant. 

Eq.\eqref{eq:gam-zero} shows a $\log(1/\epsilon)$ divergence, is the classic ``chord-length'' expression for the EE  
of a single interval in a 1+1d CFT on a circle\cite{calabrese_entanglement_2004}. The scaling in Eq.\eqref{eq:gam-zero} was numerically observed for 
a massless Dirac fermion on a thin torus in Ref.~\onlinecite{chen_scaling_2014}.  
We note that when a small twist along $y$ is introduced, $\lambda>0$, it induces an order $\lambda$ mass for the zero mode that
cuts off the $\log(L_A/\epsilon)$ divergence to $\log(1/\lambda)$. 

At $\lambda=0$ but for generic $\lambda_x$, the answer for $\gamma_n$   
is expected to differ from Eq.\eqref{eq:gam-zero}. Indeed, in the case of a real $(1+1)$d boson with anti-periodic boundary conditions $\lambda_x=1/2$,
numerical calculations on the lattice\cite{Chojnacki2016} have shown deviations from Eq.\eqref{eq:gam-zero}. More recently,   
the R\'enyi entropies for a free boson CFT with R\'enyi index $n\neq 1$ were analytically computed as a function of $\lambda_x$ by Shiba \cite{shiba}, and they differ from Eq.\eqref{eq:gam-zero}. See the  Added Note after the Conclusions. 
 

Finally, before moving to the massive case, we want to emphasize a subtle difference between our analytical and numerical calculation. For the free boson model 
in Eq.\eqref{H_boson}, we expressed the Hamiltonian in the continuum, hence we have a sum over a set of discrete and yet infinite number of Fourier modes in Eq.\eqref{sum_1d}. Alternatively, we can consider a lattice version of the free boson Hamiltonian defined on the torus, 
given in Eq.\eqref{H_discrete}. 
The effective mass for each $1+1$d chain then changes to $\sqrt{2-2\cos(k_y)+m^2}$ (where the lattice spacing is set unity),
and we only need to sum over a finite number of momenta. In the limit of large lattices that we consider, our results
agree with the analytical calculations performed in the continuum. In the thin torus limit,
a detailed comparison between the analytical and lattice results is presented in Fig.~\ref{fig:cons_term}, where good agreement is found.

\subsubsection{Massive case}

We now consider the effect of adding a mass to the theory, leading to a $m^2\phi^2$ term in the Lagrangian.
We note that for periodic boundary conditions ($\lambda=0$), a finite mass needs to be 
introduced to cure the divergence of $\gamma_n$, Eq.\eqref{gamma_2d_z_1}.   
Therefore we need to calculate this infinite sum 
\begin{align}
  g(\lambda,mL_y)=\sum_{p=-\infty}^{\infty}\log\left[(mL_y)^2+(2\pi)^2(p+\lambda)^2\right]
\end{align}
This can be regularized by using the Epstein zeta function regularization method (see Appendix \ref{epstein}), and we have
\begin{align} \label{2d_g_mass}
g(\lambda,mL_y)=\log\left[2\cosh(mL_y)-2\cos(2\pi\lambda)\right]
\end{align}
This bare result was also computed in Ref.~[\onlinecite{Whitsitt-2016}] and Ref.~[\onlinecite{Arias2015}]. 
Naively, the universal EE would then be proportional to this quantity. However, we notice a strange property: Eq.\eqref{2d_g_mass}
diverges linearly in the large mass limit, $g(\lambda,mL_y\gg1)= mL_y+\dotsb$. This is at odds with the intuition that the
infinite mass fixed point, which describes a state without spatial entanglement, should have $\gamma=0$.
This divergence points to a deeper problem with our naive procedure to extract the universal EE by subtracting the area law contribution. Indeed, when the theory is away from the massless fixed point, the additional length scale $1/m$ renders 
the subtraction procedure ambiguous. A similar situation occurs when considering the EE of a region $A$ 
in infinite flat space instead of a torus. For example when $A$ is a perfect disk living in $\mathbb R^2$, 
the universal subleading term in the EE is often called $-F$. A statement called the $F$-theorem has been shown 
for relativistic theories:\cite{casini-huerta-2012} $F$ decreases monotonically along an RG flow. If we thus have a UV CFT that flows 
into an IR one, the following inequality will hold $F_{\rm UV} > F_{\rm IR}$. Now, for a free massive boson, 
a naive computation of $F$ by subtraction of the area law leads to linear divergence $F\sim -mR$ at large mass.
Here, $R$ is the radius of the disk. This is exactly analogous to the divergence that we have seen in Eq.\eqref{2d_g_mass}.
However, the divergence of $F$ is at odds with the $F$-theorem since the infinite mass fixed point is trivial, and has $F=0$.
The cure for the disk is known: one needs to consider a \emph{renormalized} EE, $\mathcal F(R)=-S+R\,\tfrac{\partial S}{\partial R}$, 
where $S$ is the full EE (computed using any given regulator).\cite{casini-huerta-2012,Liu:2012eea}      
At conformal fixed points, $\mathcal F=F$ agrees with the
expected CFT value, while $\mathcal F(R)$ decreases monotonically along an RG flow linking two fixed points.    

\begin{figure}[hbt]
\centering
\includegraphics[width=.7\textwidth]{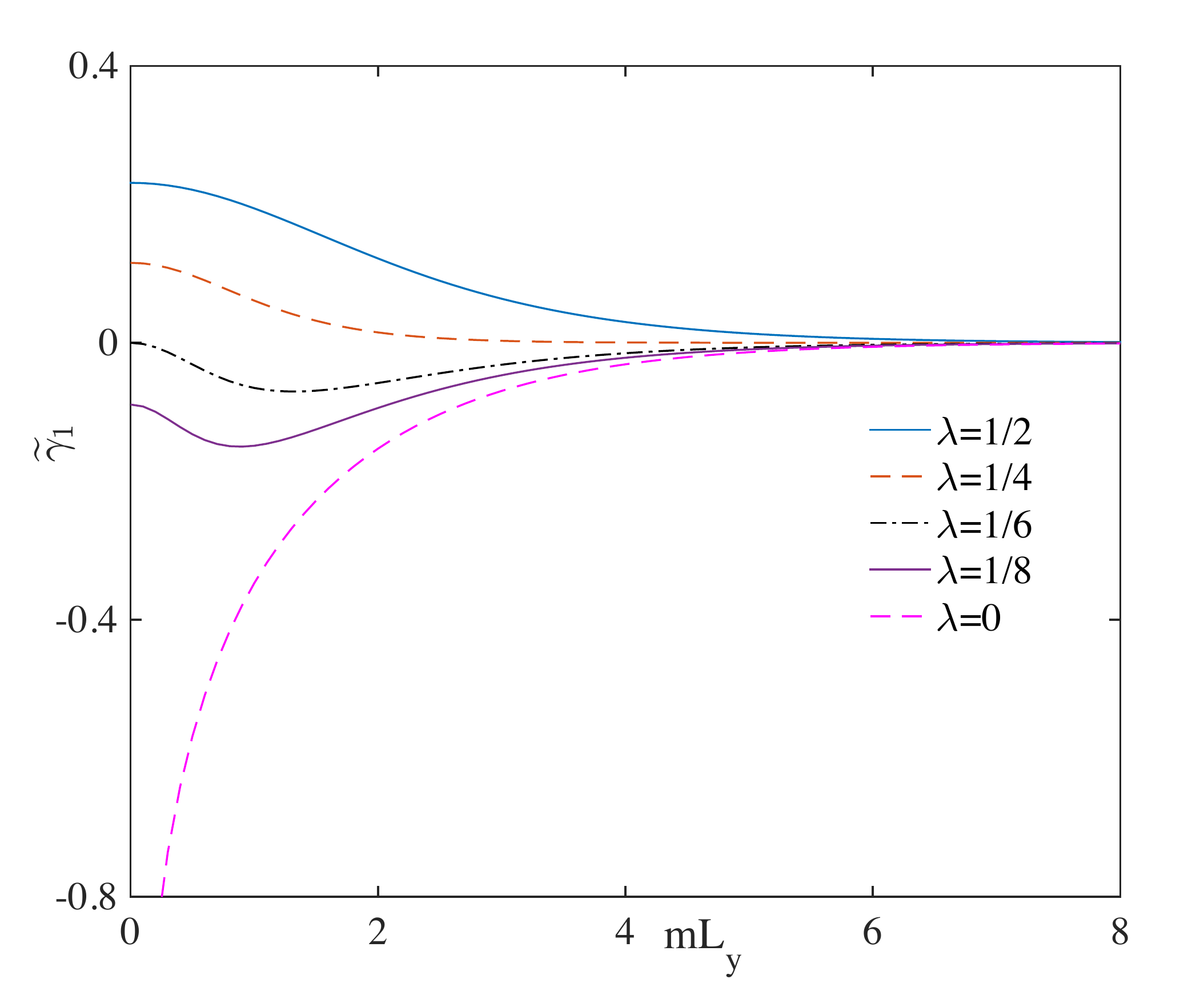}
\caption{(Color online) The renormalized EE $\widetilde\gamma_1$ for the complex boson (or two Dirac fermions), 
Eq.\eqref{ree_2d},  as a function of $mL_y$ for various twists $\lambda$. 
For periodic boundary conditions, $\lambda\!=\! 0$,
there is a divergence $-\log(mL_y)$ as $mL_y\to 0$. This agrees with the fact that $\gamma$ has a $\log$ dependence on the cutoff $\epsilon$ when $m=\lambda=0$,
see \eqref{eq:gam-zero}.
} 
\label{cons_massive}
\end{figure}

Going back to our situation on the torus, we can define a renormalized $\gamma$ as follows:\cite{Bueno-prep}         
\begin{align} \label{REE}
  2\widetilde\gamma(L_y) = -S + L_y\frac{\partial S}{\partial L_y}\,,
\end{align}
where $S$ is here evaluated in the thin torus limit $b\to \infty$.
(This can be naturally generalized beyond the thin torus limit to all $u,b$.\cite{Bueno-prep})
Eq.~(\ref{REE}) is the one-to-one analogue of the renormalized disk EE used in the $F$-theorem, discussed above. 
In contrast to the $F$-theorem however, there is no proof that this quantity is the ``natural''
one to consider, in the sense of being an RG monotone, say. Our motivation for using Eq.\eqref{REE} 
for defining the EE away from a fixed point is its simplicity, and its connection to the disk EE prescription (which is
also non-unique when $A$ is not a disk). 
In the case of the massive complex boson, we thus find:
\begin{align}  \label{ree_2d} 
2\widetilde\gamma_n(mL_y) &= \frac{1}{6}\left(1+\frac{1}{n}\right)\left[ g(\lambda,mL_y)-L_y\frac{\partial g(\lambda, mL_y)}{\partial L_y}\right] \\
&= \frac{1}{6}\left(1+\frac{1}{n}\right)\left\{ \log[2\cosh(mL_y)-2\cos(2\pi\lambda)] - \frac{mL_y \sinh(mL_y)}{\cosh(mL_y)-\cos(2\pi\lambda)} \right\} \notag
\end{align} 
In the small and large $t=mL_y$ limits, the renormalized EE becomes
\begin{align}
  \widetilde\gamma_1(t\to 0) &= \gamma_1 - \frac{t^2}{48\sin^2(\pi\lambda)} + \frac{t^4 (2+ \cos(2\pi\lambda))}{384 \sin^4(\pi\lambda)}+O(t^6) \\
  \widetilde\gamma_1(t\to\infty) &= -\frac{\cos(2\pi\lambda)}{6}\; t e^{-t} + \dotsb \label{large-m}
\end{align}
respectively. We see that the leading correction at small $t$ is always negative, i.e.\ $\widetilde\gamma_n$ 
initially decreases under the RG flow, irrespective of $\lambda$. 
Interestingly, a similar result was found in the context of strongly interacting holographic CFTs at $n=1$.\cite{Bueno-prep} 
It would be interesting 
to investigate the RG flow of $\widetilde\gamma$ in the vicinity 
of more general UV fixed points (weak detuning regime), 
in the same spirit as was done for the EE of a disk.\cite{Faulkner-2015} 
Second, we observe that $\widetilde{\gamma}_n$ decays to zero exponentially fast in the deep IR limit $t\to\infty$, Eq.\eqref{large-m}, as shown in Fig.~\ref{cons_massive}. 
However, as shown in that figure, it is not a monotonically decreasing function for all values of the twist $\lambda$.
Indeed, Eq.~\eqref{large-m} shows that $\widetilde\gamma_n$ becomes strictly increasing at large $t$ 
when $0<\lambda<1/4$.    
The non-monotonicity could have been anticipated because when $m=0$, $\gamma_n$ can be strictly negative for certain values of $\lambda$, and we 
know that at the trivial infinite mass fixed point it will vanish, $\widetilde{\gamma}_n=0$.  
Interestingly, at $\lambda=1/6$, $\widetilde{\gamma}_n$ is zero at both fixed points, $mL_y=0$ and $\infty$. 
However, it is a non-trivial function along the RG flow that interpolates between these two fixed points, as shown in Fig.~\ref{cons_massive}.

\section{$z=2$ free bosons \& fermions with a twist}  
\label{2d_z_2}

In this section we study the torus EE of two non-relativistic gapless systems: the non-interacting boson with
dynamical exponent $z=2$, and the fermionic quadratic band touching (which also has $z=2$). 

\subsection{$z=2$ free bosons}

The Hamiltonian for free boson field $\phi$ with dynamical scaling exponent $z = 2$ is
\begin{align}
  H=\int d^2x \, \frac{1}{2} \Big( \Pi^2+    \big[{\bm \nabla}^2\phi\big]^2 \Big) \, ,
  \label{eq:QLM-H}
\end{align}
where $\Pi$ is its conjugate canonical momentum. When $\phi$ is compact (i.e., $\phi\equiv \phi+2\pi R$), 
this model is  quantum Lifshitz model and describes the quantum dimer model at critical point.\cite{Ardonne-2004}
There has been a lot of references discussing EE in this model.\cite{fradkin_entanglement_2006,stephan_shannon_2009,hsu_universal_2009,Hsu-2010,oshikawa_boundary_2010,stephan_renyi_2012,stephan_entanglement_2013,Zhou2016}  Here we study the entanglement of the 
non-compact version, focusing on the subleading universal term on the torus, when there is a twist in the $y$ direction. 
The ground state wavefunction of Eq.\eqref{eq:QLM-H} has a simple and elegant form 
\begin{align}
|\psi \rangle = \frac{1}{\sqrt{Z}}\int [d\phi] e^{-\frac{1}{2}\mathcal S[\phi]} |\phi \rangle\,.
\label{eq:Psi-QLM}
\end{align}
Here $\mathcal{Z}$ is the partition function of the free boson CFT in a two-dimensional Euclidean spacetime,  
and $S[\phi]$ is the corresponding Euclidean action, 
\begin{align}
  Z = \int [d\phi] e^{-\mathcal S[\phi]}, \qquad \mathcal S[\phi] = \frac{1}{2} \int d^2x \, \left({\bm \nabla}\phi \right)^2
 \label{eq:Psi-QLM2}
\end{align} 

For the above wave function, if $\phi$ is non-compact, by using the replica trick directly on the wave function, we have \cite{fradkin_entanglement_2006}
\begin{align}
  \Tr\rho_A^n=\left(\frac{Z(A)Z(B)}{Z(A\cup B)}\right)^{n-1}
\end{align} 
where $Z(A)$ and $Z(B)$ are the free boson partition functions on region $A$ and $B$, respectively, with Dirichlet boundary 
condition $\phi=0$ on the boundary. $Z(A\cup B)$ is the boson partition function on the entire space, with the same boundary conditions
as those imposed on the 2+1 dimensional theory. 
The R\'enyi entanglement entropy is then
\begin{align}
S_n=-\log\frac{Z(A)Z(B)}{Z(A\cup B)}~,
\label{form}
\end{align}
which is independent of the R\'enyi index $n$. Below we consider the entire system $A\cup B$ to be an open cylinder and torus, respectively.

\subsubsection{Two-cylinder entropy on a cylinder}

\begin{figure}
\centering
\vspace*{0cm}
\includegraphics[width=.5\textwidth]{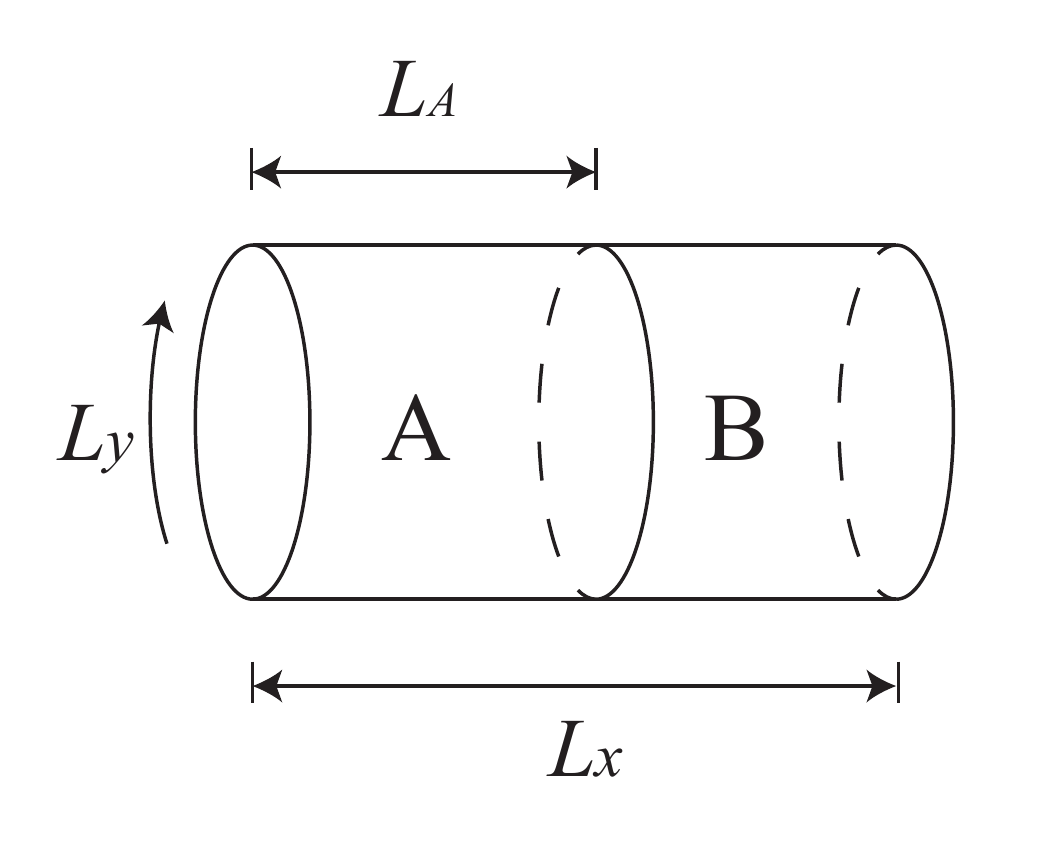}
\caption{Bipartition of a cylindrical geometry. We impose Dirichlet boundary conditions $\phi=0$ at both ends.
The universal part of the EE is $-\mathcal J$, which is generally distinct from the analogous quantity defined on a torus, $-J$. }
\label{cylinder}
\end{figure}

We first study the groundstate on the open cylinder and calculate the two-cylinder EE with twisted boundary condition in the $y$ direction, which is also denoted by $\lambda$ and is defined through $\phi(x,y+L_y)=e^{2\pi i\lambda}\phi(x,y)$. We impose Dirichlet boundary condition with $\phi=0$ on both ends of the entire cylinder $A\cup B$. As shown in Fig.\thinspace\ref{cylinder}, $Z(A)$, $Z(B)$ and $Z(A\cup B)$ are all partition functions defined on cylinder.  For the real boson, the only possible twist boundary condition corresponds to $\lambda=1/2$. Under this twist, the partition function of the cylinder (after regularization) is equal to 
\begin{align}
Z=\sqrt{2}\sqrt{\frac{\eta\!\left(-\frac{1}{2\tau}\right)}{\theta_2\!\left(-\frac{1}{2\tau}\right)}}=\sqrt{2}\sqrt{\frac{\eta(2\tau)}{\theta_4(2\tau)}}
\end{align}
where $\tau=iL_x/L_y$ is the modular parameter and $u=L_A/L_x$. The explicit form of  $\eta(\tau)$, $\theta_2(\tau)$ and $\theta_4(\tau)$ are shown in Appendix \ref{ap:useful}.  
According to Eq.\eqref{form}, the subleading correction 
$-\mathcal J_n(u)$ of EE is equal to 
\begin{align}
  \mathcal J_n(u)=\frac{1}{2}\log\left( \frac{2\eta(2u\tau) \eta(2(1-u)\tau)\theta_4(2\tau)}{\eta(2\tau)\theta_4(2u\tau)\theta_4(2(1-u)\tau)} \right)
\end{align}
In the thin cylinder limit $\tau\to\infty$, if the ratio $u$ takes a finite value, the cylinder partition function $Z\to \sqrt{2}$ and therefore $J_n(u)=\log(\sqrt{2}\times \sqrt{2}/\sqrt{2})=\log(\sqrt{2})$.  

For a complex boson, $\lambda\in (0,1)$. The partition function on the cylinder (after regularization) under a general twist is equal to 
\begin{align}
Z&=e^{\pi i(\lambda-\frac{1}{2})} {(1-e^{-2\pi i\lambda})}{\frac{\eta(-\frac{1}{2\tau})}{
\thg{\frac1 2}{\lambda-\frac1 2}(-\frac{1}{2\tau})}}\nonumber\\ 
&={(1-e^{-2\pi i\lambda})}{\frac{\eta(2\tau)}{ \thg{\lambda-\tfrac{1}{2}}{-\tfrac{1}{2}}(2\tau)}}
\end{align} 
 The explicit form of the $\thg{\alpha}{\beta}(\tau)$ function is given in Eq.\eqref{theta_twist}. 
According to Eq.\eqref{form}, the universal EE contribution for the two-cylinder bi-partition of the open
cylinder in case of a complex boson is
 \begin{align}
   \mathcal J_n(u)=\log\left(\frac{\eta(2u\tau)\eta(2(1-u)\tau) \thg{\lambda-\frac1 2}{-\frac1 2}(2\tau) 
   }  {\eta(2\tau) \thg{\lambda-\frac1 2}{-\frac1 2}(2u\tau)
   \thg{\lambda-\frac1 2}{-\frac1 2}(2(1-u)\tau)}\right)
   +\log\left(1-e^{-2\pi i\lambda}\right) 
\label{J_cylinder}
 \end{align} 
$\mathcal J_n(u)$ here is written in terms of theta and Dedekind $\eta$ functions, and is fairly non-trivial. 
In the following sections, we examine various limits of $\mathcal J_n$.\\

\paragraph{Thin slice limit:}
The partition function for cylinder A of the complex boson is equal to
\begin{align}
Z(A)=e^{\frac{\pi}{12ub}}\prod_{n=1}^{\infty}\frac{1}{(1-e^{2\pi i\lambda}e^{-\frac{\pi n}{ub}})(1-e^{-2\pi i\lambda}e^{-\frac{\pi n}{ub}})}
\end{align}
If we keep the ratio $b=L_x/L_y$ finite in the thin slice limit $u\to 0$, $Z(A)\to e^{\frac{\pi}{12ub}}$. 
The remaining contribution to $S_n$ is $\log [Z(B)/Z(A\cup B)]\approx 0$, and therefore we have 
\begin{align}
  \mathcal J_n(u\to 0)=\frac{\pi c}{24ub}=\frac{\pi cL_y}{24L_A}
\end{align}
where $c$ is the central charge for the classical two dimensional conformal field theory. 
For the complex boson we discuss here, it is equal to $c=2$, while for the real boson, $c=1$. 
Notice that in this limit, $\mathcal J_n(u)$ is independent of the twists in the $x$ and $y$ directions.\\

\paragraph{Thin cylinder limit:} 
We take $L_y\to 0$ at fixed $u$, which means $\tau\to i\infty$. Then the first term in Eq.\eqref{J_cylinder}
is equal to $\log(e^{\pi i(\lambda-\frac{1}{2})})$. When combined with the second term in $S_n$, 
we find that the EE saturates to a constant independent of any length scale, $\gamma_n$: 
\begin{align}  \label{gam_z2_boson}
  \gamma_n=\log\left[(1-e^{-2\pi i\lambda})e^{\pi i(\lambda-\frac{1}{2})}\right]=\log\left[2\sin(\pi\lambda)\right]
\end{align}
which is exactly the same $\gamma_n$ that appears in the universal EE of a thin \emph{torus}, Eq.\eqref{gamma-all} and
Eq.\eqref{thin-torus_z2_boson} below. However, in the torus case the EE tends to $2\gamma_n$ instead because of the presence
of two boundaries. The agreement, up to  this trivial factor of 2, between the open-cylinder and torus geometries in the $L_y\to 0$ limit follows because
the EE is insensitive to degrees of freedom far from the entanglement cut, where distances ought
to be compared to the limiting length scale $L_y$.
We note that the $\lambda$-dependence of $\gamma_n$ for the $z=2$ boson is the same as for the $2+1$ dimensional free boson and Dirac fermion CFTs,
as given in Eq.\eqref{gamma_2d_z_1}, up to an overall prefactor. 

The $\mathcal J$-function in Eq.\eqref{J_cylinder} can be compared with that for the quantum Lifshitz model, which takes this form on the open cylinder:\cite{Zhou2016}
\begin{align}
\mathcal J_1^{\rm QLM}(u)=\log \left| \frac{\eta(2\tau)}{\eta(2\tau u)\eta(2\tau(1-u))} \right|
-\frac{1}{2}\log[ 2u(1-u)|\tau|] + W(\tau,R)
\end{align}
where $W(\tau,R)$ is a $u$-independent term coming from the zero mode sector for the compact boson with compactification radius $R$. The boson is here assumed to be periodic along $y$.
In the thin slice limit, $\mathcal J_1^{\rm QLM}(u\to 0)=\pi/(24ub)$, which is the same as $\mathcal J_1$ of the non-compact $z=2$ boson,
Eq.\eqref{J_cylinder}. In the thin cylinder limit, $L_y\to 0$, $\mathcal J_1^{\rm QLM}$ tends to a shape-independent constant, 
$-\log (\sqrt{4\pi }R)+\frac{1}{2}$, which only depends on $R$.\cite{hsu_universal_2009,Hsu-2010,oshikawa_boundary_2010}  
As $R\to\infty$, one recovers the non-compact boson answer. Indeed $\mathcal{J}_1^{\rm QLM} \to -\infty$ is the 
same as $\gamma_1$ in the non-compact case in the limit $\lambda \to 0$, Eq.\eqref{gam_z2_boson}.     

\subsubsection{Two-cylinder entropy on a torus}  
If the total system is on a torus, there are two boundaries between $A$ and $B$. By using the partition function 
of the torus $Z(A\cup B)$, Eq.\eqref{part_torus}, we can write down the $J$-function for two-cylinder EE defined on the torus, 
\begin{align}
J_n(u)=&\log\left(\frac{\eta(2u\tau)\eta(2(1-u)\tau)
\thg{\lambda-\frac{1}{2}}{-\lambda_x+\frac{1}{2}}(\tau)
\thg{-\lambda+\frac{1}{2}}{-\lambda_x+\frac{1}{2}}(\tau) 
}  {\eta^2(\tau)
\thg{\lambda-\frac{1}{2}}{-\frac{1}{2}}(2u\tau)
\thg{\lambda-\frac{1}{2}}{-\frac{1}{2}}\left(2(1-u)\tau\right)
}\right) 
+2\log(1-e^{-2\pi i\lambda}) 
\end{align}
where $\lambda_x$ and $\lambda$ represent the twists in $x$ and $y$ directions, respectively. 

\begin{figure}
  \centering
  \vspace*{-0cm}
  \includegraphics[width=.7\textwidth]{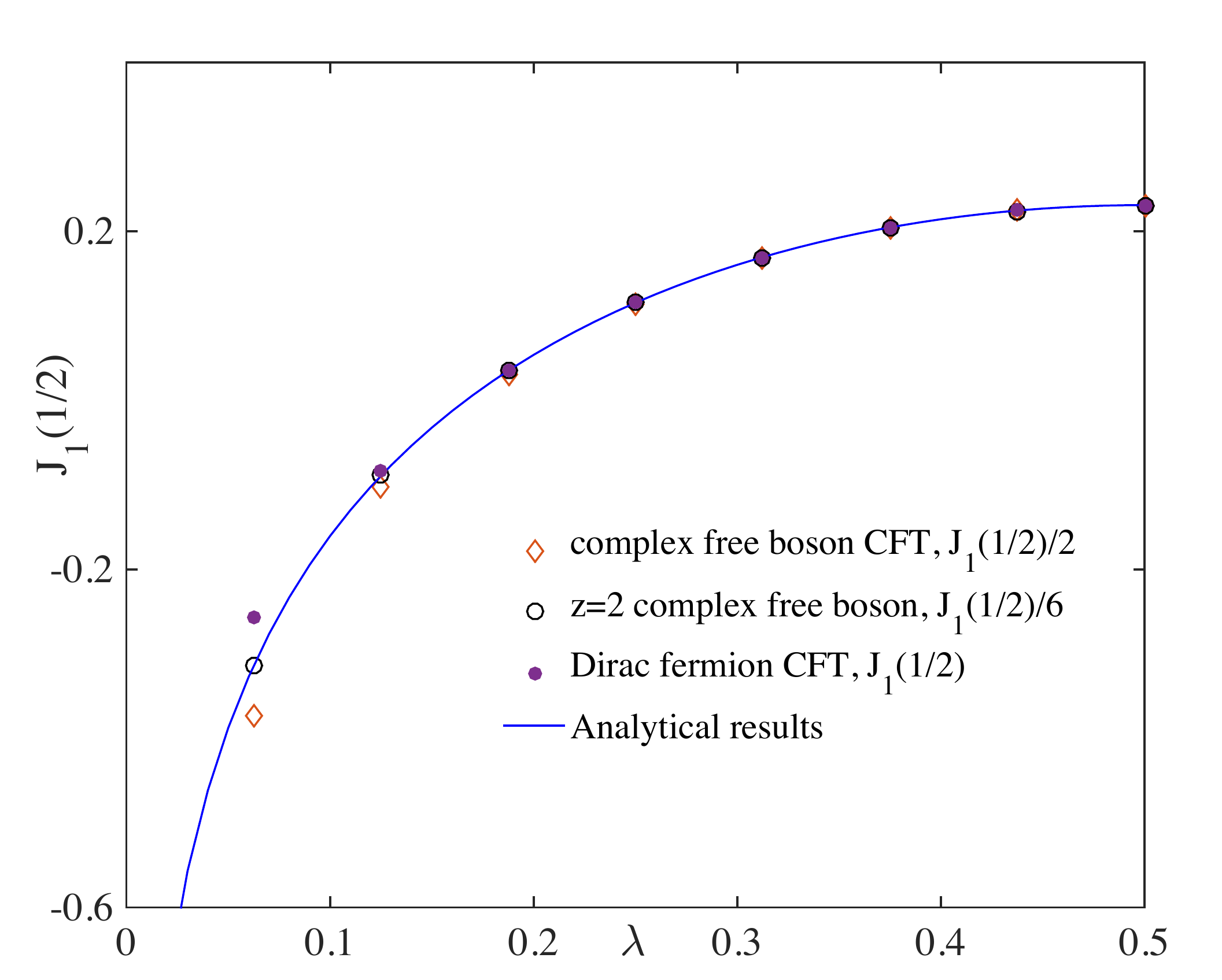}
  \caption{(Color online) The two cylinder EE on the torus $J_1(1/2)$ at $b=6$ versus the twist parameter along the 
    $y$-direction, $\lambda$. For such a large value of $b$, $J_1(1/2)$ approaches $2\gamma_1$, Eq.~\eqref{gamma-all}.  
    We have rescaled the data so that it collapses to one curve: $\tfrac{1}{3}\log[2\sin(\pi\lambda)]$. 
    For the bosonic models, each point is obtained by taking $L_y$ from $200$ to $400$ and subtracting the area law term. 
    For the Dirac fermion, each point is obtained by taking $L_y$ from $200$ to $320$ and subtracting the area law term. 
    The $z=2$ free boson model matches best with the analytical expression. The mismatch for free boson and Dirac fermion CFTs at small $\lambda$ is due to 
    finite size effects.} 
  \label{gamma_2d}
\end{figure}

\paragraph{Thin torus limit:} In the thin torus limit, $J_n(u)$ approaches a constant $2\gamma_n$ when $\lambda>0$, 
which is independent of $\lambda_x$, and is twice the value of the corresponding coefficient for a bipartition of an infinite cylinder
into two equal halves:
\begin{align} \label{thin-torus_z2_boson}
  2\gamma_n=2\log\left[(1-e^{-2\pi i\lambda})e^{\pi i(\lambda-\frac{1}{2})}\right]=2\log\left( 2\sin(\pi\lambda) \right).
\end{align}
The overall factor of 2 arises because of the two boundaries between $A$ and $B$.  
Therefore, the $z=2$ massless free boson, free boson CFT and Dirac fermion CFT all have the same twist dependence but 
with different overall prefactors.   
We numerically obtain $2\gamma_1$ by calculating $J_1(1/2)$ on the lattice at large $b$ for these  
three models and the results are shown in Fig.~\ref{gamma_2d}. For $\lambda$ close to $1/2$, the numerical results agree perfectly
with the analytical expression for $2\gamma_1$. As $\lambda$ decreases, the Dirac fermion and free boson CFT start to show deviations 
from $2\gamma_1$. This is because as $\lambda$ decreases, the $1+1$-dimensional modes around $k_y=0$ have smaller effective masses, 
and larger $b$ and $L_y$ are therefore required to obtain better agreement.

\paragraph{Wide torus limit:} We now consider the limit opposite to the thin torus one, namely the wide torus: $L_y/L_x\to\infty$ (Fig.\ref{fig:shapes} (c)). 
If $u$ is finite, we have   
\begin{align}
\log Z(A)+\log Z(B)=\frac{\pi c}{24}\frac{1}{bu(1-u)}
\end{align}
This result is obtained by using the derivation in Eq.\eqref{part_cyl_s}. In contrast, the partition function $Z(A\cup B)$ on the entire
torus will contribute a $u$-independent term. 
In the limit $b\to 0$, it is
\begin{align}
\log Z(A\cup B)=\frac{\pi c}{6b}
\end{align}
Therefore we have
\begin{align} \label{z2-wide-torus}
J_n(u)=\frac{\pi c}{24b}\frac{1}{u(1-u)}-\frac{\pi c}{6b}
\end{align}
where for a real bosonic field, $c=1$, while $c=2$ for the complex boson. 
Eq.\eqref{z2-wide-torus} can be re-written as 
\begin{align}
  J_n(u)=\frac{2\pi}{b}a_n(\theta)\,, \qquad \theta=2\pi u\,,
  \label{J_theta}
\end{align}
where
\begin{align} \label{lif-corner}
a_n(\theta) = \frac{c}{12}\frac{ (\theta -\pi)^2}{\theta (2\pi - \theta)}
\end{align}
We have used the natural mapping between the cylinder's normalized length $u$, and the \emph{angle} $\theta=2\pi u$. 
Actually, $a_n(\theta)$ is the coefficient of the logarithmic term in the EE for the $z=2$ boson 
that arises in the presence of a sharp corner of opening angle $\theta$ in the entangling surface.\cite{fradkin_entanglement_2006}  
This is because the infinite cylinder can be mapped to the infinite plane through a conformal transformation. 
The transformation maps an infinite strip of width $u$ living on the cylinder to a wedge of angle $\theta=2\pi u$ on the plane. 
To make this clear, we can also directly calculate the corner correction of EE for $z=2$ boson model. 
Since $\log Z$ for a wedge with opening angle $\theta$ is equal to \cite{cardy_finite-size_1988}
\begin{align}
\log Z=-\frac{c\theta}{24\pi}\left(1-\frac{\pi^2}{\theta^2}\right)\log( L/\epsilon),
\end{align}
Following Ref.~\onlinecite{fradkin_entanglement_2006}, one readily finds that the contribution to the entanglement entropies follows from adding the corner contribution to the free energies of regions $A$ (with angle $\theta$) and $B$ (with angle $2\pi-\theta$), leading to the result that 
 the R\'enyi entropies have a subleading logarithmic contribution associated with a corner of angle $\theta$ of the form 
\begin{align}
S_n = B_n L - a_n(\theta) \log(L/\epsilon) +\dotsb
\end{align}
where the corner coefficient for the $z=2$ boson is given in Eq.\eqref{lif-corner}. We emphasize that this direct relation
between the wide-torus limit of $J_n(u)$ and the corner coefficient $a_n(\theta)$ arises here because the ground state
is invariant under the infinite group of two-dimensional conformal transformations.\cite{Ardonne-2004} 
This is not the case for general critical theories like the gapless boson and Dirac fermion CFTs.

The case $\lambda=0$ must be treated separately, just as was done for CFTs. 
We leave such analysis for the future.

\subsection{Fermionic quadratic band touching} 

The fermionic quadratic band touching (QBT) model describes free fermions
with a quadratic energy dispersion ($z=2$). The low-energy Hamiltonian for the QBT takes the form 
\begin{align} 
H=\int \frac{d^2k}{(2\pi)^2} \Psi^{\dagger} (\bm k) \begin{pmatrix} k_x^2-k_y^2 & -2ik_xk_y \\ 2ik_xk_y& -k_x^2+k_y^2 \end{pmatrix}\Psi(\bm k), 
\label{H_QBT} 
\end{align}  
where we have defined the two-component spinor $\Psi(\bm k)=(\psi_1(\bm k),\psi_2(\bm k))^T$. 
We have set the band-curvature scale $M$ to unity, $k_ik_j/M\to k_i k_j$, as it will play no role in
our discussion.
In contrast to the Dirac fermion, this model corresponds to a critical point not a critical phase,  
and has a finite density of states at the band touching point.\cite{sun-2009} 

The QBT model naturally satisfies the area law due to the absence of an extended Fermi surface. 
On the square torus, $b=1$, the $u$-dependence of the correction $J_1(u)-J_1(1/2)$ has been numerically studied in Ref.\ \onlinecite{chen_scaling_2014}.
Unlike for the $z=2$ free boson model, it is not known whether the groundstate for QBT model is connected to 
a two dimensional CFT, and therefore the analytical methods used to study the $z=2$ free boson 
cannot be applied.  
\begin{figure}
  \centering
  \vspace*{-0cm}
  \includegraphics[width=.7\textwidth]{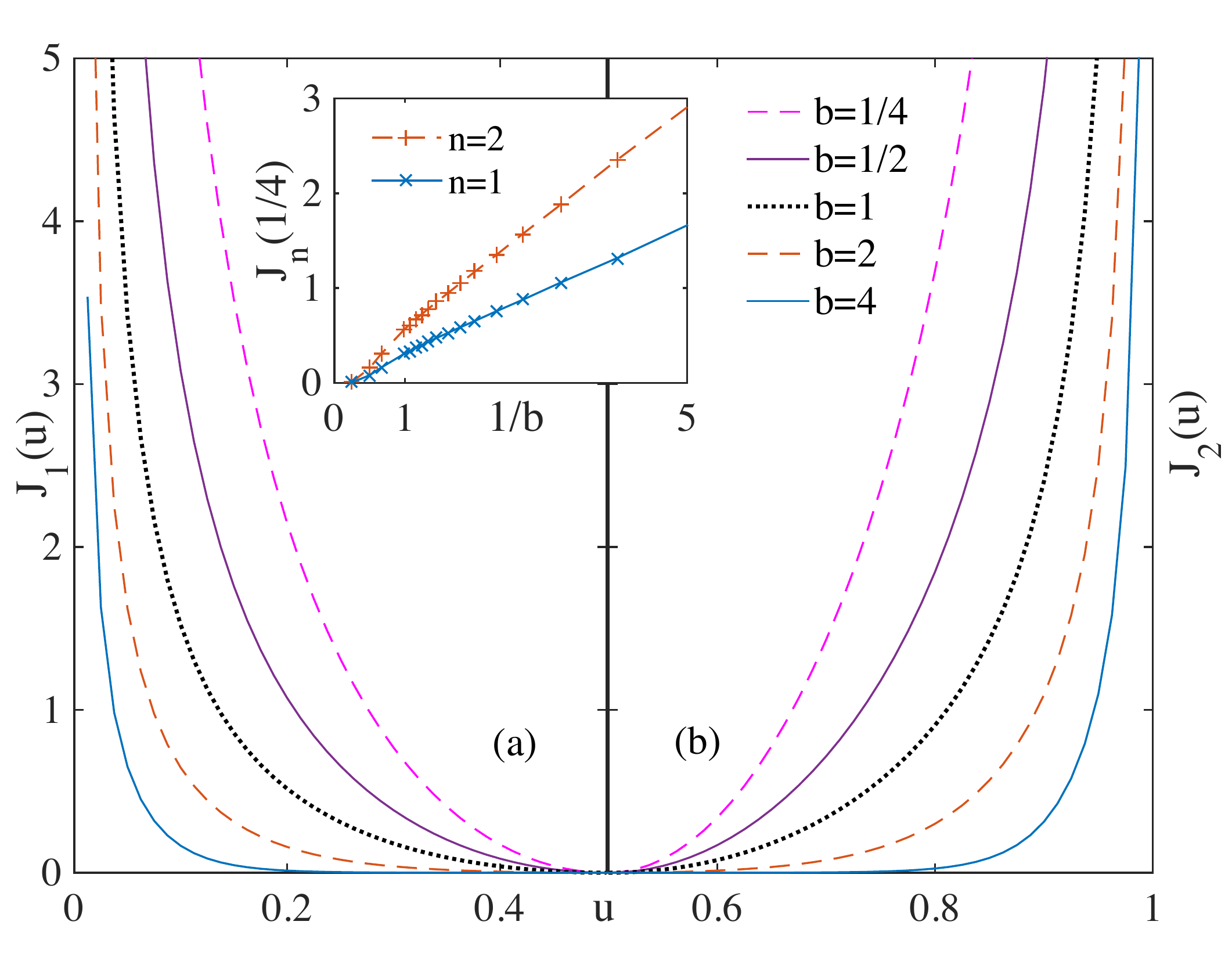}
  \caption{(Color online) The universal contribution $J_1(u)$ of the QBT model for various aspect ratios $b=L_x/L_y$, with anti-periodic boundary condition in the $y$-direction and periodic boundary condition in $x$. $L_x=320$. The inset shows $J_n(1/4)$ 
    as a function of $1/b$, where the line is a guide to the eye.} 
  \label{QBT_J}
\end{figure} 

We numerically study the $u$ and $b$ dependence of $J_n(u;b)$ and find that the different R\'enyi indices lead to similar behavior. Therefore we will focus only on $n=1,2$.
Different from the other three models we have studied in $2+1$ dimensions, we find that when $A$ covers half the torus, $J_n(1/2; b)$  
vanishes for arbitrary boundary conditions along $x$ and $y$:
\begin{align} \label{QBT-surprise}
  J_n(1/2; b) = 0\,.
\end{align}
This is true for \emph{all} aspect ratios $b$. In particular, taking the thin torus limit $b\to\infty$, Eq.~\eqref{QBT-surprise}
implies that $\gamma_n = 0$. 
This behavior at $u=1/2$ distinguishes the QBT from all the critical theories that have been studied so far,
and it would be interesting to understand the physical reasons underlying Eq. \eqref{QBT-surprise}. 
As shown in Fig.~\ref{QBT_J}, as $u\to 1/2$, all the curves indeed approach zero. 

In the wide torus limit, $b\to 0$, we previously found that the $z= 2$ boson has $\lim_{b\to 0} b J_n(1/2;b)=0$,
which also holds for the QBT by virtue of Eq.~\eqref{QBT-surprise}. In contrast, the non-interacting boson 
and Dirac fermion CFTs have $J_n(1/2)\sim 1/b$ as $b\to 0$. This is also the case for certain CFTs described by the AdS/CFT
holographic correspondence.\cite{chen_scaling_2014}  
For $u\neq 1/2$, when $b\ll 1$, the QBT has $J_n(u)\sim 1/b$, in accordance with our general argument. In the inset of Fig.~\ref{QBT_J}, we plot $J_1(1/4)$ and $J_2(1/4)$ as a function of $1/b$ to illustrate this fact.  

In the thin slice slice limit $u\to 0$, taken at fixed $b$, we obtain the expected divergence\cite{chen_scaling_2014,krempa2016} $J_n(u) =\kappa_n/(bu)=\kappa_n L_y/L_A$,
with
\begin{align} \label{qbt-kap}
  \kappa_1 = 0.182\,, \qquad\quad \kappa_2 = 0.263\,.
\end{align}
We first note that $\kappa_2>\kappa_1$, which is distinct from the behavior exhibited by the free boson and Dirac fermion CFTs, 
where $\kappa_n$ decreases with $n$ for $n=1,2,3,4,\infty$.\cite{Bueno_twist}
The values Eq.~\eqref{qbt-kap} can be compared with those of the free Dirac fermion:\cite{Casini_rev,Bueno_twist} $\kappa_1^{\rm Dirac}=0.0722$ and $\kappa_2^{\rm Dirac}= 0.0472338$,
which are  $2.5$ and $5.6$ smaller than for the QBT, respectively. This is in line with the heuristic expectation that $\kappa$ is a measure of the gapless degrees of freedom
since a QBT can be split into a pair of Dirac fermions at different momenta. The splitting can be accomplished by adding a term of the form $A\sigma_z$ to the Hamiltonian Eq.~\eqref{H_QBT}. 

\section{Free relativistic theories in $3+1$ dimensions} 
\label{3d_z_1} 

We now consider the free relativistic complex boson, and 4-component Dirac fermion in three spatial dimensions.  
Let us consider their groundstate on the three dimensional torus $L_x\times L_y\times L_z$, and take region $A$ to be
the cylinder $L_A\times L_y\times L_z$, as shown in Fig.~\ref{3d_torus}. 
The entanglement entropies will take the following form: 
\begin{align}
  S_n = \alpha_n \frac{2L_y L_z}{\epsilon^2} - J_n(u;b_1,b_2) +\dotsb
\end{align}
with the two aspect ratios being $b_1=L_x/L_y$ and $b_2=L_x/L_z$. 
We note that there is no logarithm because the entangling surface is not curved. $J_n$ will also depend on the boundary conditions
along the three non-contractible cycles. 
The function $J_1(u;b_1,b_2)-J_1(1/2;b_1,b_2)$ was studied in Ref.~\onlinecite{krempa2016} for the free complex boson at fixed boundary conditions
and $b_1=b_2$. 
Here we focus on the \emph{thin torus limit} of the full $J_n(u;b_1,b_2)$, 
where $L_y$ and $L_z$ tend toward zero at fixed $L_x$ and $L_A$: 
\begin{align}
  \lim_{L_y,L_z\to 0} J_n(u;b_1,b_2) = 2\gamma_n^{3d}(r)\,, \qquad r= L_y/L_z\,. 
\end{align} 
We shall study the general twist dependence.
In the thin torus limit, the R\'enyi entropies can be obtained by Fourier transforming along $y$ and $z$,
which maps the problem to a sum over massive 1d bosons/fermions. Since $L_y,L_z\ll L_A,L_x$, we have
\begin{align}  \label{3dlog-decomposition}
  S_n=-\sum_{k_x,k_y}\frac{1}{6} \left(1+\frac{1}{n}\right) \log[(k_y^2+k_z^2+m^2)\epsilon^2 ] 
\end{align}
where we have used the $1+1$-dimensional expression for the EE,\cite{calabrese_entanglement_2009} setting the Virasoro central charge to $c=2$ since we work 
with a complex boson or a 4-component Dirac fermion.
The momenta are quantized as follows: $k_y=2\pi (n_1+\lambda_1)/L_y$, $k_z=2\pi (n_2+\lambda_2)/L_z$, $n_1,\ n_2\in\mathbb{Z}$ and $\lambda_1,\lambda_2\in (0,1)$. 
The case $\lambda_1=\lambda_2=0$ requires special care, and we treat it separately below. 
To extract the universal subleading correction to the EE, 
we need to evaluate the double series
\begin{align}
g(mL_y;\lambda_i,r)=\sum_{n_1,n_2=-\infty}^{\infty}\log\left[\left(\frac{mL_y}{2\pi}\right)^2+(n_1+\lambda_1)^2+r^2(n_2+\lambda_2)^2\right]
\label{3d_sum}
\end{align}
where $r=L_y/L_z$. We consider the massless $m=0$ and massive $m\neq 0$ cases separately.

\subsection{Massless case} 

\begin{figure}[hbt]
\centering
\includegraphics[width=.7\textwidth]{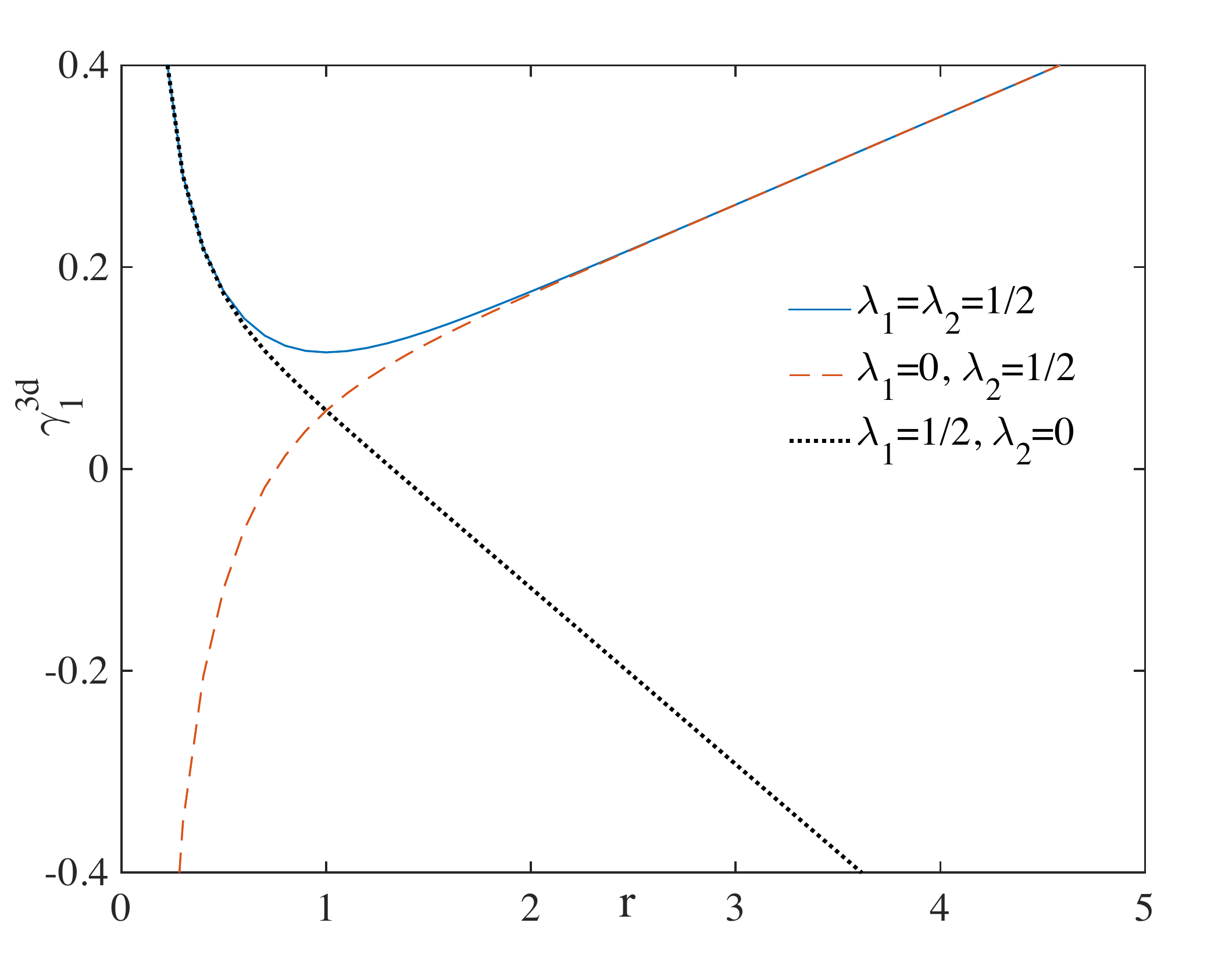}
\caption{(Color online) The function $\gamma^{3d}_1$ for the universal term of the von Neumann EE of a complex
massless boson / 4-component Dirac fermion in 3+1 dimensions for a semi-infinite partition of an infinite cylinder. 
$\gamma_1^{3d}$, given in Eq.\eqref{massless},  
is plotted as a function of the torus aspect ratio $r=L_y/L_z$, and for various values of the twists $\lambda_1$ and $\lambda_2$.} 
\label{cons_3d_massless}
\end{figure} 

For the conformal state, $m=0$, the double series of Eq.\eqref{3d_sum} becomes
\begin{align}
  g(0;\lambda_i, r)=\sum_{n_1,n_2=-\infty}^{\infty}\log\left[(n_1+\lambda_1)^2+r^2(n_2+\lambda_2)^2\right]
\end{align}
As shown in Appendix \ref{epstein_2d}, this series is given, after regularization, by the expression 
\begin{align} \label{g_3d}
 g(0; \lambda_i, r ) = \log\left( \frac{ \thg{\lambda_2- \frac{1}{2}}{\lambda_1-\frac{1}{2}}(\tau) }{\eta(\tau)} \cdot 
 \frac{\thg{\lambda_2- \frac{1}{2}}{-\lambda_1 +\frac{1}{2}}(\tau) }{ \eta(\tau)} \right)
\end{align}
where $\tau= ir=i L_y/L_z$ is the modular parameter of the toroidal boundary. 
We have assumed $\lambda_1,\lambda_2 > 0$, i.e.\ we exclude simultaneous periodic boundary conditions along $y$ and $z$.
The $\lambda_1=\lambda_2=0$ case requires special care, as we discuss below. 
Therefore, in the massless case, the subleading universal term to the EE is (for $\lambda_1,\lambda_2>0$)
\begin{align} 
  2\gamma_n^{3d} = \frac{1}{6}\left(1+\frac{1}{n}\right) \log\left( \frac{ \thg{\lambda_2- \frac{1}{2}}{\lambda_1-\frac{1}{2}}(\tau) }{\eta(\tau)} \cdot 
 \frac{\thg{\lambda_2- \frac{1}{2}}{-\lambda_1 +\frac{1}{2}}(\tau) }{ \eta(\tau)} \right) 
  \label{massless} 
\end{align}
In the special case $\lambda_1=\lambda_2$ and $r=1$, $\gamma_n^{3d}$ was derived in Ref.\ \onlinecite{Metlitski2009} for the complex boson, 
where they used the replica trick method, and expressed the result in terms of the first Jacobi theta function.
Eq.\eqref{massless} reduces to their answer when $\lambda_1 \!=\!\lambda_2$, as can be seen by using
the relation between $\theta{\alpha \brack \beta}(\tau)$ and the first Jacobi theta function $\theta_1$, 
Eq.\eqref{elliptic}. We should emphasize that our result in  Eq.\eqref{massless} is valid for 
arbitrary values of the twists $\lambda_i$ (as long as $\lambda_1,\lambda_2 > 0$) and $r$, and that it also holds for the Dirac fermion. 

Our setup is symmetric under the interchange of the $y$ and $z$ directions, which means that
$J_n(1/2)$ is invariant under the simultaneous exchanges $r\leftrightarrow 1/r$ and $\lambda_1\leftrightarrow\lambda_2$. 
This symmetry can be seen by taking $\tau\to -1/\tau$, $\lambda_1\to\lambda_2$, $\lambda_2\to\lambda_1$ in Eq.\eqref{g_3d},
and using the relations of Eq.~\eqref{S_theta}, $\eta(-1/\tau)=\sqrt{-i\tau}\eta(\tau)$, and $\theta{\alpha\brack -\beta}(-1/\tau)=\sqrt{-i\tau}e^{2\pi\alpha\beta}\theta{\beta\brack -\alpha}(\tau)$.

Since the closed form answer Eq.~\eqref{massless} is somewhat opaque, we find it useful to examine special limits
where $\gamma^{\rm 3d}$ simplifies.
For instance when the boundary becomes very elongated in one direction, say $r=L_y/L_z \to \infty$, we find
\begin{align} \label{large-r}
  2\gamma_n^{3d}=-\left(1+\frac{1}{n}\right)\frac{2\pi r}{3}\left[ \frac{1}{2}\left(\lambda_2-\frac{1}{2}\right)^2-\frac{1}{24} \right]
\end{align}
Thus, $\gamma_n^{3d}$ diverges linearly with $r$ and only depends on the twist 
in the (short) $z$ direction, $\lambda_2$. This asymptotic behavior  is seen in Fig.~\ref{cons_3d_massless}.
Conversely, in the opposite limit, $r\to 0$, we find instead
\begin{align} \label{small-r}
2\gamma_n^{3d}=-\left(1+\frac{1}{n} \right)\frac{2\pi }{3r}\left[ \frac{1}{2}\left(\lambda_1-\frac{1}{2}\right)^2-\frac{1}{24} \right]
\end{align}
In this case, it only depends on the twist in $y$ direction and is independent of $\lambda_2$. We note that Eq.~(\ref{small-r}) can be obtain from Eq.~(\ref{large-r}) by using the symmetry that interchanges the $y,z$ directions. The $1/r$ divergence is 
shown in Fig.~\ref{cons_3d_massless}.  

Fig.~\ref{cons_3d_massless} shows the full result for $\gamma^{3d}_1$ as a function of $r$ and different values of $\lambda_1$ and $\lambda_2$. Notice that  $\gamma_1$ can be both negative and positive depending on the value of the twists $\lambda_1$ and $\lambda_2$. We see that there is a 
apparent divergence as $\lambda_1,\lambda_2\to 0$, which we now turn to
by examining the periodic case $\lambda_1=\lambda_2=0$.  

{\bf Periodic case, $\lambda_1=\lambda_2=0$:} We can no longer use Eq.\eqref{3dlog-decomposition} because of the 
zero mode $k_y=k_z=0$. The zero mode will contribute a term dependent on $u=L_A/L_x$ and $L_x/\epsilon$ where $\epsilon$ is 
the short distance cutoff, as well as on the
twist in the $x$-direction $\lambda_x$. This is exactly as for the 2+1d CFTs, and is a consequence of the momentum decomposition of the EE in the free CFTs:
$S_n=\sum_{k_y,k_z} S_n^{1d}(k_y,k_z)$, where the transverse momenta determine the effective masses of the 1d modes; we have omitted the dependence on the twist parameters.  
When $\lambda_1=\lambda_2=0$, the 1d mode with $k_y=k_z=0$ has zero mass and will dominate the EE.
It is exactly the same zero mode encountered in 2+1d, which implies that
\begin{align}
  \gamma_n^{3d}\big|_{\lambda_1=\lambda_2=0} = \gamma_n^{2d}\big|_{\lambda = 0} + \dotsb
\end{align}
where the dots denote a constant independent of $L_x,\epsilon,\lambda_x$. For periodic boundary conditions along $x$, $\lambda_x=0$, we have, 
see Eq.\eqref{eq:gam-zero},
\begin{align} \label{eq:gam-zero2}
  \gamma_n^{3d} \big|_{\lambda_1=\lambda_2=0} = -\frac16\left(1+\frac{1}{n}\right) 
\log \left(\frac{L_x}{\pi\epsilon} \sin\left(\frac{\pi L_A}{L_x}\right)\right) + \dotsb
\end{align} 
For $\lambda_x>0$, the R\'enyi entropies for a free boson CFT in 1+1d were analytically computed as a function of $\lambda_x$ by Shiba \cite{shiba}, and these
have to be used instead of \eqref{eq:gam-zero2}. For the Dirac fermion, the answer is not known at present. 
Finally, we note that by turning on a small $\lambda_1,\lambda_2>0$, the apparent logarithmic divergence $\log(1/ \epsilon )$ will be cutoff, leading to the 
$\log(\lambda_{1,2})$ scaling seen above.

\subsection{Massive  case}

\begin{figure}[hbt]
  \centering
  \includegraphics[width=.6\textwidth]{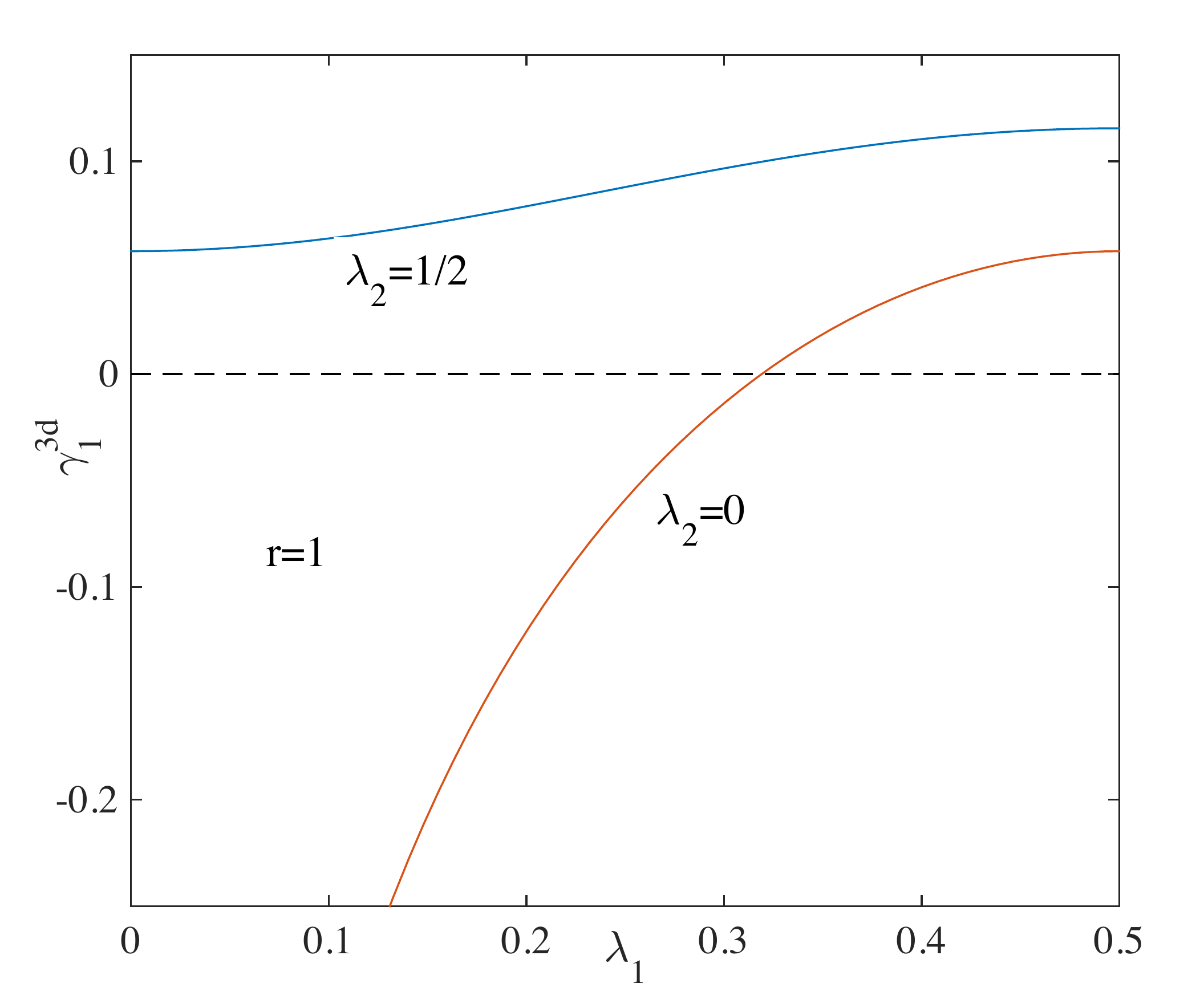}
  \caption{ (Color online) The universal term $\gamma^{3d}_1$ of the von Neumann EE for a massless complex
    boson / 4-component Dirac fermion in 3+1 dimensions, Eq.~\eqref{massless}, plotted as a function of $\lambda_1$ 
    for a torus with aspect ratio $r=1$.
    The top (bottom) curve corresponds to $\lambda_2=1/2$ $(0)$. For $\lambda_2=0$, $\gamma^{3d}_1$ vanishes at $\lambda_1\simeq 0.319$.} 
  \label{cons_3d_lambda}
\end{figure}

 We turn on a finite mass $m>0$. We work in the thin torus limit $L_y,L_z\ll L_A, L_x$,
and study $\gamma^{3d}$ as function of $mL_y$ and $mL_z$, and the twists. This means that we keep $m L_{y,z}$ order unity. 
This is equivalent to working on the semi-infinite cylinder with $L_x$ and $L_A$ infinite (where we would get $\gamma^{3d}$ instead of $2\gamma^{3d}$).
To evaluate the double sum over modes, Eq.\eqref{3d_sum}, we can first sum over $n_1$. According to Eq.\eqref{Epstein_reg}, we have 
\begin{align}
  g(mL_y; \lambda_i,r)= g_1(mL_y;\lambda_i,r)+g_2(mL_y;\lambda_i, r)
\end{align} 
where $g_1$ is given by
\begin{align}
g_1(mL_y;\lambda_i,r ) =2\pi \sum_{n_2\in\mathbb{Z}}\sqrt{\left(\frac{mL_y}{2\pi}\right)^2+r^2(n_2+\lambda_2)^2}~,
\end{align}
and $g_2$ by 
\begin{align}
g_2(mL_y;\lambda_i, r)=\sum_{n_2\in\mathbb{Z}}\log\left[ (1-e^{2\pi i\lambda_1-2\sqrt{t}\pi})(1-e^{-2\pi i\lambda_1-2\sqrt{t}\pi}) \right]
\end{align}
with $t=(\tfrac{mL_y}{2\pi})^2+r^2(n_2+\lambda_2)^2$.
$g_2$ is finite but $g_1$ is divergent and needs to be further regularized. Using the derivation in Appendix \ref{epstein}, we can separate the divergent and finite contributions to $g_1$,
\begin{align}
g_1=-\frac{(mL_y)^2}{4\pi r}\Gamma(-1)-\frac{(mL_y)^2}{\pi r}\sum_{p\neq 0}e^{2\pi i p\lambda_2}\frac{K_{1}(z)}{z}
\end{align}
where $z=\frac{mL_y}{r}|p|$, $\Gamma(-1)=\infty$ and $K_{\nu}(z)$ is the modified Bessel function of the second kind.

\begin{figure}[hbt]
\centering
 \subfigure[]{\label{fig:a05} \includegraphics[width=.48\textwidth]{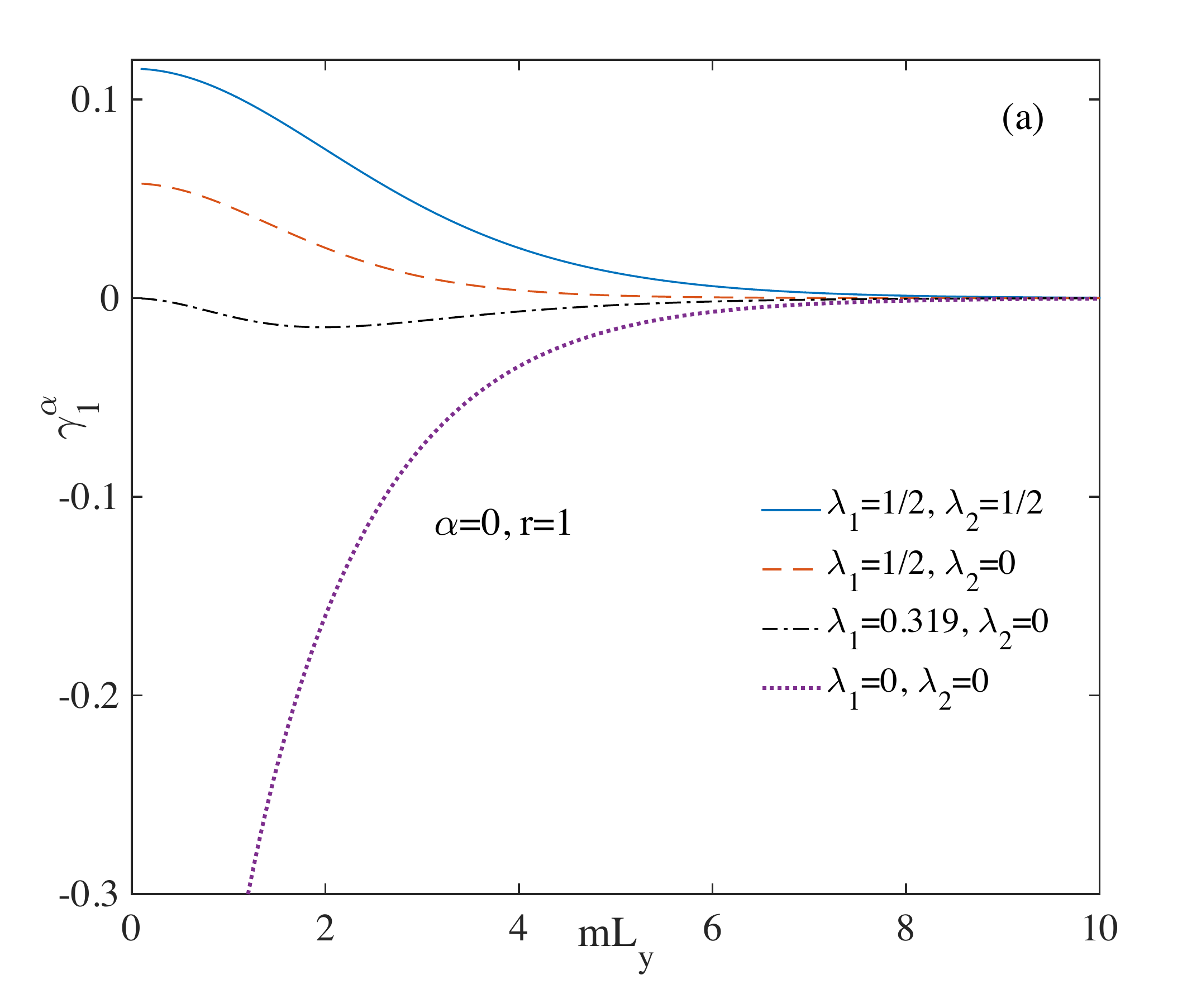}}
 \subfigure[]{\label{fig:a0} \includegraphics[width=.48\textwidth]{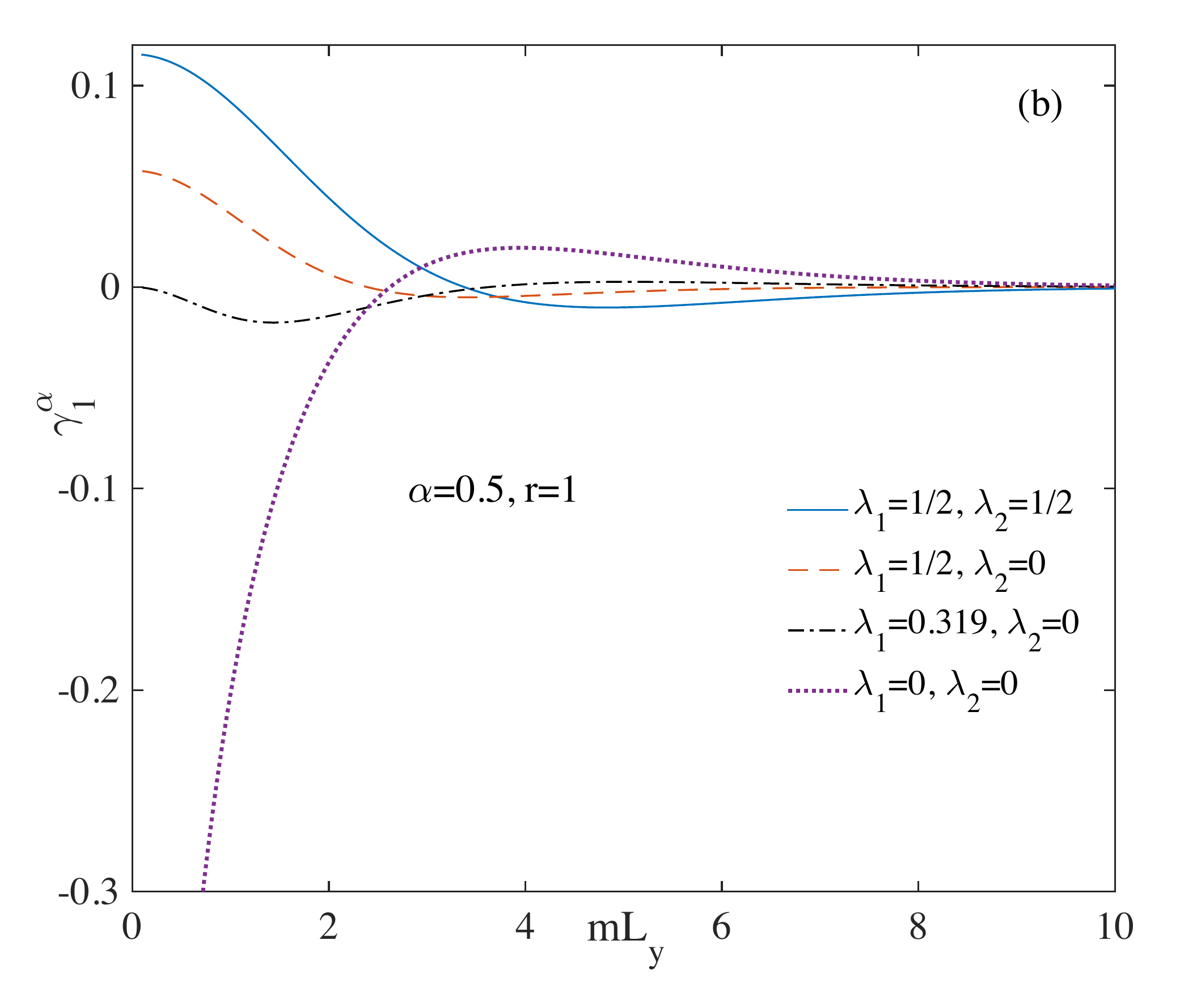}}   
\caption{(Color online) The renormalized von Neumann EE $\gamma^{(\alpha)}_1$, 
(a) for $\alpha=0$, and (b) $\alpha=1/2$,
given in Eq.\eqref{mass-gam-3d}, for a complex boson / 4-component Dirac fermion of mass $m$ in 3+1 dimensions. 
It is a scaling function of $mL_y$, and is shown for four choices of the twists
$\lambda_{1,2}$, and fixed torus aspect ratio $r=1$.} 
\label{cons_3d_massive}
\end{figure}

Just as was the case in 2+1 dimensions, we cannot simply subtract the area law in order to obtain
a sensible answer for $\gamma_n^{3d}$ when the mass is finite. Working in the thin torus limit, we consider the one-parameter family 
of renormalized EEs: \cite{Bueno-prep}
\begin{align} \label{REE-3d}
2\gamma_n^{(\alpha)}(mL_y;r)=&\frac{1}{2}\left[\alpha L_y^2\frac{\partial^2S_n(L_y)}{\partial L_y^2}+(1-\alpha)L_y\frac{S_n(L_y)}{\partial L_y}-2S_n(L_y)\right]
\end{align}
which is the most general form that ensures a cancellation of the area law, 
is linear in $S_n$, and includes terms up to second order in $L_y$-derivatives. The case linear in derivatives, $\alpha=0$,
is the natural analogue of the 2+1 dimensional renormalized EE used above, Eq.~\eqref{REE}.
The aspect ratio $r=L_y/L_z$ is kept as a fixed constant to ensure that the entangling 
geometry does not change shape as we probe the entanglement 
along the RG flow parametrized by $mL_y$. At a scale-invariant fixed point, Eq.~(\ref{REE-3d}) reduces 
to $2\gamma_n^{3d}(r)$.
In $\gamma_n^{(\alpha)}$, the divergent term in $g_1$ will cancel and we have
\begin{align} \label{mass-gam-3d}
  2\gamma^{(\alpha)}_n(mL_y)=- \frac{1}{12}\left(1+\frac{1}{n}\right) \left[ \alpha L_y^2\frac{\partial^2(\tilde{g}_1+g_2)}{\partial L_y^2}+(1-\alpha)L_y\frac{\partial(\tilde{g}_1+g_2)}{\partial L_y}-2(\tilde{g}_1+g_2) \right]
\end{align}
where $\tilde{g}_1$ is the finite part of $g_1$. 
We evaluate this expression numerically and plot it in Fig.~\ref{cons_3d_massive} for $\alpha=0, 1/2$ and $r=1$. 
In the deep IR limit, $m L_y\to\infty$, $\gamma^{(\alpha)}_1$ eventually approaches zero. 
The fact that $\gamma=0$ at the trivial gapped IR fixed point has been discussed in Ref.\ \onlinecite{Grover2011}.
In most cases, it is not 
monotonically decreasing with $mL_y$. Actually, in the massless limit, $\gamma_n^{(\alpha)}(mL_y)$ reduces to Eq.\eqref{massless} by construction.
Clearly, it is not possible to define $\gamma^{(\alpha)}_n$ to be a monotonic decreasing function since at $mL_y=0$ (IR), as shown in 
Fig.~\ref{cons_3d_lambda}, already in the massless limit $\gamma_1^{3d}$ can take either positive or negative values as a function of 
the twists $\lambda_1$ and $\lambda_2$. 
In the special case $\lambda_1 \simeq 0.319$ and $\lambda_2=0$, $\gamma^{3d}$ is strictly equal to zero at the massless point, 
but we find that no matter how we tune $\alpha$, $\gamma^{(\alpha)}_1(mL_y)$ becomes finite at $mL_y>0$, which means
that it cannot be monotonic since it also approaches zero as $mL_y\to \infty$. 
Although $\gamma^{(\alpha)}_1$ is generally not a monotonic function along the RG flow, it is nevertheless a simple and non-trivial entanglement 
 measure that is universal (independent of the UV cutoff). 


\section{Conclusions} 
\label{conclusion}

In this paper, we investigated the subleading correction term $J_n(u)$ of two-cylinder EE for several scale invariant free systems in both $2+1$ dimensions and $3+1$ dimensions. We numerically studied $J_n(u,b)$ 
under various twisted boundary conditions in $2+1$ dimensions. In the thin torus limit, we found that $J_n(u,b)$ converges to some constant $2\gamma_n$ which only depends on the twist and is independent of any length scale of the system. We further evaluated this constant analytically for Dirac fermion CFT, free boson CFT and massless $z=2$ free boson system and find that they take the same form, as a function of the twist, up to a prefactor. This is in contrast to the fermionic quadratic band touching model, where $\gamma_n$ is always equal to zero. We further extended our calculations to $3+1$ dimensions and obtained the analytical expression for $\gamma^{3d}_n$ in the thin torus limit for both free boson CFT and Dirac fermion CFT with twisted boundary conditions.

The finite constant terms in the EE contain non-trivial and universal information on the properties of scale-invariant field theories. In addition to providing new insights about these fixed-point theories they may also constrain the allowed RG flows between two fixed points after we deform away from the UV theory.  Indeed a long-standing problem is to find possible generalizations of the celebrated Zamolodchikov's $c$-theorem of 1+1-dimensional field theories which showed that under the action of a relevant perturbation the relativistic RG flow always connects two fixed points with decreasing values of the conformal central charge $c$, and that this change is described by a monotonically decreasing $c$-function along the RG flow.\cite{Zamolodchikov-1986}

One of the problems faced in this endeavor is to find a quantity that would play the role similar to that of the central charge of the trace anomaly of the energy momentum tensor.\cite{Cardy-1988} This quantity is unique only in 1+1-dimensional CFTs and already in 3+1 dimensions there are two such ``central charges'', named $c$ and $a$ respectively. In odd space-time dimensions there are no such quantities associated with the energy-momentum tensor. The discovery that  in 1+1 dimensional CFTs the central charge $c$ determines the coefficient of the logarithmic term of the EE
as well as the demonstration that EE provides a \emph{new} RG monotone 
has turned the problem of finding generalizations of Zamolodchikov's theorem to that of finding similar behavior for EE in higher dimensional theories.

In 3+1 dimensional relativistic theories  an $a$-theorem for $a$ central charge has
been established.\cite{Komargodski-2011} Here $a$ enters in the logarithmic corrections to the area law term of the EE for a spherical entangling surfaces. 
In this case the actual $c$-function has not been proven to be related to an EE along the RG-flow,
instead a certain integrated cross section, related to a four point function of the
trace of the stress tensor, plays this role.
The analog of this result in 2+1 dimensional relativistic CFTs  is the $F$-theorem\cite{Klebanov-2011,jafferis2011,casini-huerta-2012} associated with the finite universal 
corrections to the area law for circular entangling surfaces and of an associated function defined with respect to the EE which 
decreases monotonically along RG flows. There are further distinctions between the $a$-theorem and the $F$-theorem in that the latter is not obviously related to properties of any known \emph{local} operator and $F$ can count topological degrees of freedom\cite{kitaev_topological_2006,levin2006} while these do not show up in $a$.

In this paper we have investigated the RG flow of
similar finite universal quantities which exist in the EE for cylindrical cuts of 2+1-dimensional theories on a torus. The goal here was to search in the possible parameter space (including twisted boundary conditions and various geometries) for an RG monotone. 
The quantities we studied in this paper, just like $F$,  have a more non-local origin as they are also able  to capture the contributions of non-local degrees of freedom, such as those in topological field theories\cite{Dong-2008} and in scale-invariant theories with compactified fields.\cite{hsu_universal_2009,metlitski2011entanglement,agon2014disk,Zhou2016}  However, we only ended up ruling out various reasonable possibilities since the quantities we study do not obey the sought after monotonicity property. 
Early attempts to generalize the  Zamolodchikov $c$-theorem using the thermodynamic entropy density (which is a natural physical quantity to use to estimate the number of degrees of freedom)  also failed to exhibit a monotonically decreasing behavior.\cite{Castro_Neto-1993,Sachdev-1993}

Moving forward one could now expand the search for RG monotones to other geometries and entangling cuts. There are several reasons why we think this is a useful pursuit. The existence of $c$-functions from a relativistic point of view is on fairly sound footing in $1+1, 2+1$ and $3+1$ dimensions albeit without a unified picture in these different dimensions. Perhaps some new entanglement entropy quantity will provide such a unification. Similarly for non-relativistic theories positive results on $c$-functions are non-existent and filling this gap would likely have deep consequences. \\

{\bf Added Note:} After submission of this paper, a preprint\cite{shiba} of Noburo Shiba appeared, giving the 
analytical calculation of the entanglement R\'enyi entropies in a free boson CFT on a circle in 1+1d as a function of the
twist $\lambda_x$. This is a lower dimensional analogue of our present analysis in 2+1d and 3+1d.
There exists a close connection between our results and the 1+1d result of Shiba in the thin torus limit, as explained in
Sections \ref{2d_z_1} and \ref{3d_z_1}.

\begin{acknowledgements}
We thank Pablo Bueno, Lauren Hayward Sierens, Roger Melko, Subir Sachdev, Noburo Shiba, Alex Thomson, and Seth Whitsitt. This work was supported in part by the National Science Foundation through the grant DMR 1408713 at the University of Illinois (EF). XC was supported by a postdoctoral fellowship from the the Gordon and Betty Moore Foundation, under the EPiQS initiative, Grant GBMF-4304, at the Kavli Institute for Theoretical Physics. 
WWK was supported by a postdoctoral fellowship and a Discovery Grant from NSERC, by a Canada Research Chair, and by
MURI grant W911NF-14-1-0003 from ARO. 
WWK further acknowledges the hospitality of the Aspen Center for Physics, where 
part of this work was done, and which is supported by the National Science Foundation through the grant PHY-1066293. TF is supported by the DARPA YFA program, contract D15AP00108.

\end{acknowledgements}

\appendix
\section{Useful functions and identities} \label{ap:useful}

We define the following function
\begin{align}
\thg{\alpha}{\beta}(\tau) =
\eta(\tau)\,e^{2\pi i \alpha\beta}q^{\frac{\alpha^2}{2}-\frac{1}{24}}\prod_{n=1}^{\infty}
\left(1+q^{n+\alpha-\frac{1}{2}}e^{2\pi i\beta}\right)\left(1+q^{n-\alpha-\frac{1}{2}}e^{-2\pi i\beta}\right)
\label{theta_twist}
\end{align}
where $q=e^{2\pi i\tau}$, and $\eta(\tau)$ is the Dedekind eta function:
\begin{align}
\eta(\tau)=q^{\frac{1}{24}}\prod_{n=1}^{\infty}(1-q^n)\,.
\end{align}
Under the S transformation, $\tau\to-1/\tau$, we have
\begin{align} \label{S_theta} 
  \theta{\alpha \brack \beta}(-1/\tau)=\sqrt{-i\tau} e^{2\pi i\alpha\beta} \theta{\beta\brack -\alpha}(\tau)
\end{align}
We also define
\begin{align}
&\theta_2(\tau)\equiv\thg{\frac{1}{2}}{0}(\tau)=2q^{1/8}\prod_{n=1}^{\infty}(1-q^n)(1+q^n)^2,\nonumber\\
&\theta_4(\tau)\equiv\thg{0}{\frac{1}{2}}(\tau)=\prod_{n=1}^{\infty}(1-q^n)(1-q^{n-\frac{1}{2}})^2
\end{align}
which are obtained from the full theta Jacobi functions $\theta_\nu(z,\tau)$ by setting $z=0$.
We note that $\thg{\alpha}{\beta}(\tau)$ can be expressed in terms of the theta Jacobi function $\theta_1(z,\tau)$,
as shown in Eq.\eqref{elliptic}. 

\section{Various zeta functions}
The details of these zeta functions can be found in Ref.\ \onlinecite{elizalde_ten_2012}.
\subsection{Hurwitz zeta function}
\label{hurwitz}
Hurwitz zeta function is defined in this way,
\begin{align}
\zeta(s,\lambda)=\sum_{n=0}^{\infty}\frac{1}{(n+\lambda)^s}
\end{align}
This is Hurwitz zeta function and it satisfies
\begin{align}
\zeta^{\prime}(0,\lambda)=-\sum_{n=0}^{\infty}\log(n+\lambda)=\log\Gamma(\lambda)-\frac{1}{2}\log (2\pi)
\end{align}
Therefore we can regularize the infinite sum
\begin{align}
g(\lambda)&=\sum_{n=-\infty}^{\infty}\log |n+\lambda|=\log\prod_{n=0}^{\infty}(n+\lambda)\prod_{n=0}^{\infty}\left[n+(1-\lambda)\right]\nonumber\\
&=-\zeta^{\prime}(0,\lambda)-\zeta^{\prime}(0,1-\lambda)\nonumber\\
&=-\log\Gamma(\lambda)+\frac{1}{2}\log(2\pi)-\log\Gamma(1-\lambda)+\frac{1}{2}\log(2\pi)\nonumber\\
&=-\log\left[\Gamma(\lambda)\Gamma(1-\lambda)\right]+\log(2\pi)\nonumber\\
&=\log(2\sin(\pi\lambda))
\label{inf_sum_g}
\end{align}
where in the last step, we used 
$\Gamma(\lambda)\Gamma(1-\lambda)=\pi/\sin(\pi\lambda)$.

\subsection{Epstein zeta function in $d=1$}
\label{epstein}
Define
\begin{align}
f(s,\lambda,c,\alpha)&=\sum_{n=-\infty}^{\infty}\left[\frac{1}{c+\alpha(n+\lambda)^2}\right]^s\nonumber\\
&=\frac{1}{\Gamma(s)}\sum_{-\infty}^{\infty}\int_0^{\infty}dt\ t^{s-1}e^{-t[c+\alpha(n+\lambda)^2]}\nonumber\\
&=\frac{\sqrt{\pi/\alpha}}{\Gamma(s)}\sum_{-\infty}^{\infty}\int_0^{\infty}dt\ t^{s-\frac{3}{2}}e^{-\frac{\pi^2}{\alpha t}n^2-ct+2\pi i\lambda n}
\label{epstein_f}
\end{align}
where we use
\begin{align}
&\Gamma(s)=\int_0^{\infty}dx\ x^{s-1}e^{-x}\nonumber\\
&\sum_{n\in\mathbb{Z}}e^{-\pi an^2+bn}=\frac{1}{\sqrt{a}}\sum_{k\in\mathbb{Z}}e^{-\frac{\pi}{a}(k+\frac{b}{2\pi i})^2}
\end{align}
with $a=\frac{\pi}{\alpha t}$ and $b=2\pi i\lambda$.

The $n=0$ part gives 
\begin{align}
f^{(n=0)}(s,\lambda,c,\alpha)=\frac{\sqrt{\pi/\alpha}}{c^{s-\frac{1}{2}}}\frac{\Gamma(s-\frac{1}{2})}{\Gamma(s)}
\end{align}
For $n\neq 0$, we have
\begin{align}
f^{n\neq 0}(s,\lambda,c,\alpha)&=\frac{\sqrt{\pi/\alpha}}{\Gamma(s)}\sum_{n\neq 0}e^{2\pi i\lambda n}\int_0^{\infty}\frac{dt}{c^{s-\frac{1}{2}}}t^{s-\frac{3}{2}}e^{-t-\frac{c\pi^2 n^2}{\alpha t}}\nonumber\\
&=\frac{2c^{-s+\frac{1}{2}}\sqrt{\pi/\alpha}}{\Gamma(s)}\sum_{n\neq 0}e^{2\pi i\lambda n}\left(\frac{z}{2}\right)^{s-\frac{1}{2}}K_v(z)
\end{align}
where $K_v(z)$ is the modified Bessel function of the second kind
\begin{align}
K_v(z)=\frac{1}{2}\left(\frac{z}{2}\right)^v\int_0^{\infty}dt\ e^{-t-\frac{z^2}{4t}}t^{-v-1}
\end{align}
where $v=-s+\frac{1}{2}$ and $z=2\sqrt{c/\alpha}\pi|n|$.
$K_{\nu}(z)=\sqrt{\frac{\pi}{2z}}e^{-z}+\ldots$. Around $s=0$, $\Gamma(s)\sim 1/s$.  We expand $f(s,\lambda,c,\alpha)$ around $s=0$,
\begin{align}
f(s\to 0,\lambda,c,\alpha)&=\frac{\Gamma(s-\frac{1}{2})}{\Gamma(s)}\frac{\sqrt{\pi}}{(c/\alpha)^{-\frac{1}{2}}}+s\sum_{n\neq 0}\frac{e^{2\pi i\lambda n}}{|n|}e^{-2\sqrt{c/\alpha}\pi|n|}\nonumber\\
&=-2\pi s (c/\alpha)^{\frac{1}{2}}-s\log\left[(1-e^{2\pi i\lambda-2\sqrt{c/\alpha}\pi})(1-e^{-2\pi i\lambda-2\sqrt{c/\alpha}\pi})\right]
\label{Epstein_reg}
\end{align}
where we use $\Gamma(-1/2)=-2\sqrt{\pi}$ and $\log (1-x)=-\sum_{k=1}^{\infty} x^k/k$.

\subsection{Epstein zeta function in $d=2$}
\label{epstein_2d}
We define the following double series
\begin{align}
g(\lambda)=\sum_{n_1,n_2\in\mathbb{Z}}\log\left[(n_1+\lambda_1)^2+r^2(n_2+\lambda_2)^2\right]
\end{align}
To regularize the above equation, we define
\begin{align}
f(s,\lambda)=\sum_{n_1,n_2\in\mathbb{Z}}\left[\frac{1}{(n_1+\lambda_1)^2+r^2(n_2+\lambda_2)^2}\right]^s
\end{align}
Here we first sum over $n_1$. According to the result in Eq.\eqref{Epstein_reg}, around $s=0$, we have
\begin{align}
f(s\to 0,\lambda)&=\sum_{n_2=-\infty}^{\infty}\frac{\Gamma(s-\frac{1}{2})}{\Gamma(s)}\frac{\sqrt{\pi}}{c^{s-\frac{1}{2}}}-\sum_{n_2=-\infty}^{\infty}s\log\left[(1-e^{2\pi i\lambda_1-2\sqrt{c}\pi})(1-e^{-2\pi i\lambda_1-2\sqrt{c}\pi})\right]
\end{align}
where $\sqrt{c}=|n_2+\lambda_2|r$. 

Hence 
\begin{align}
&f(s\to 0,\lambda)=-s\sum_{n_2=-\infty}^{\infty}\log\left[(1-e^{2\pi i\lambda_1-2r|n_2+\lambda_2|\pi})(1-e^{-2\pi i\lambda_1-2r|n_2+\lambda_2|\pi})\right]\nonumber\\
&+\frac{r\sqrt{\pi}\Gamma(s-\frac{1}{2})}{\Gamma(s)}\left[\sum_{n_2=0}^{\infty}\frac{1}{(n_2+\lambda_2)^{2s-1}}+\sum_{n_2=0}^{\infty}\frac{1}{(n_2+1-\lambda_2)^{2s-1}}\right]\nonumber\\
=&-s\sum_{n_2>0}\log\left[(1-e^{2\pi i\lambda_1-2r(n_2+\lambda_2-1)\pi})(1-e^{2\pi i\lambda_1-2r(n_2-\lambda_2)\pi})\right.\nonumber\\
&\times\left.(1-e^{-2\pi i\lambda_1-2r(n_2+\lambda_2-1)\pi})(1-e^{-2\pi i\lambda_1-2r(n_2-\lambda_2)\pi})\right]-2rs\pi\left[\zeta(2s-1,\lambda_2)+\zeta(2s-1,1-\lambda_2)\right]\nonumber\\
=&-s\sum_{n_2>0}\log\left[(1-e^{2\pi i\lambda_1-2r(n_2+\lambda_2-1)\pi})(1-e^{2\pi i\lambda_1-2r(n_2-\lambda_2)\pi})\nonumber\right.\\
&\left.\times(1-e^{-2\pi i\lambda_1-2r(n_2+\lambda_2-1)\pi})(1-e^{-2\pi i\lambda_1-2r(n_2-\lambda_2)\pi})\right]+2rs\pi (\lambda_2^2-\lambda_2+\frac{1}{6})
\end{align}
where $\zeta(s,x)$ is the Hurwitz zeta function
\begin{align}
\zeta(s,x)=\sum_{n=0}^{\infty}\frac{1}{(n+x)^s},\quad \zeta(-1,x)=-\frac{x^2-x+\frac{1}{6}}{2}
\end{align}
Therefore
\begin{align}
g(\lambda)=-f^{\prime}(s,\lambda)|_{s=0}=\log\left\{\frac{\thg{\lambda_2-\frac{1}{2}}{\lambda_1-\frac{1}{2}}(\tau)}{\eta(\tau)}
\frac{\thg{\lambda_2-\frac{1}{2}}{-\lambda_1+\frac{1}{2}}(\tau)}{\eta(\tau)}\right\}
\end{align}
where $\tau=ir$, $q=e^{2\pi i\tau}$. 
The above theta-function in the above equation is related with the first elliptic theta function in the following way,
\begin{align}
  \frac{\thg{\lambda_2-\frac{1}{2}}{\lambda_1-\frac{1}{2}}(\tau)}{\eta(\tau)}
  &=e^{2\pi i(\lambda_2-\frac{1}{2})(\lambda_1-\frac{1}{2})}e^{\pi i\tau(\lambda_2^2-\lambda_2+\frac{1}{6})}\prod_{n=1}^{\infty}(1-q^{n-1}e^{2\pi i(\lambda_1+\lambda_2\tau)})(1-q^ne^{-2\pi i(\lambda_1+\lambda_2\tau)})\nonumber\\
  &=ie^{2\pi i(\lambda_2-\frac{1}{2})(\lambda_1-\frac{1}{2})}e^{\pi i\tau\lambda_2^2+\pi i\lambda_1}\left[-ie^{\pi i z}q^{\frac{1}{12}}\prod_{n=1}^{\infty}(1-q^ne^{2\pi iz})(1-q^{n-1}e^{-2\pi iz})\right]\nonumber\\
  &=ie^{2\pi i(\lambda_2-\frac{1}{2})(\lambda_1-\frac{1}{2})}e^{\pi i\tau\lambda_2^2+\pi i\lambda_1}\frac{\theta_1(z,\tau)}{\eta(\tau)}
  \label{elliptic}
\end{align}
where $z=-(\lambda_1+\lambda_2\tau)$, and $\theta_1(z,\tau)$ is the first Jacobi theta function. 

\section{Free boson partition function on the cylinder \& torus} 
\subsection{Torus}

\subsubsection{Periodic boundary conditions in x \& y directions} 
Here we follow the method described in Chapter 10.2 of Ref.~\onlinecite{difrancesco_conformal_2012}. The partition function for non-compact free boson on the torus without the zero-mode is
\begin{align}
Z&=\int [d\varphi]\sqrt{A} \delta\left(\int d^2x \varphi\varphi_0\right) \exp\left(-\frac{1}{2}\int d^2x (\nabla\varphi)^2\right)\nonumber\\ 
&=\sqrt{A}\int \prod_n dc_n\exp\left(-\frac{1}{2}\sum_n\omega_nc_n^2\right)\nonumber\\
&=\sqrt{A}\prod_n\left(\frac{2\pi}{\omega_n}\right)^{\! 1/2} 
\end{align}
where $A=L_xL_y$ is the area of the torus, and $\varphi_0=1/\sqrt{A}$ is the normalized eigenfunction 
of the zero-mode. We have expanded the field in terms of the eigenfunctions of the Laplacian operator, $\varphi=\sum_n c_n \phi_n(x)$, with corresponding eigenvalues $\omega_n$.
We define
\begin{align}
G(s)=\sum_{n,\omega_n\neq 0}\frac{1}{\omega_n^s}
\end{align}
It satisfies
\begin{align}
\frac{d}{ds}G(s)=-\sum_n\log(\omega_n)\frac{1}{\omega_n^s}
\end{align}
Therefore we have
\begin{align}
Z(\tau)=\sqrt{A}\exp\left(\frac{1}{2}G^{\prime}(0)\right)
\end{align}
The eigenvalues $\omega_{n,m}$ are labeled by $k_x$ and $k_y$,
\begin{align}
\omega_{n,m}=k_x^2+k_y^2
\end{align}
with $k_x=\frac{2\pi n}{L_x}$ and $k_y=\frac{2\pi m}{L_y}$. Hence
\begin{align}
\left|\frac{2\pi L_x}{A}\right|^{2s}G(s)&=\sum_{(m,n)\neq (0,0)}\frac{1}{|m+n\tilde{\tau}|^{2s}}\nonumber\\
&=\sum_{m\neq 0}\frac{1}{|m|^{2s}}+\sum_{n\neq 0}\left(\sum_m\frac{1}{|m+n\tilde{\tau}|^{2s}}\right)\nonumber\\
&=2\zeta(2s)+\sum_{n\neq 0}\left(\sum_m\frac{1}{|m+n\tilde{\tau}|^{2s}}\right)
\end{align}
where $\zeta(z)$ is the Riemann $\zeta$ function, $\tilde{\tau}=-1/\tau$ and $\tau=iL_x/L_y$. Here $\tau$ is a pure imaginary number. For the second term, using the result in Sec.\ref{epstein}, when $s\to 0$, we have
\begin{align}
  \sum_{n\neq 0}\sum_m\frac{1}{|m+n\tilde{\tau}|^{2s}}&=-s\sum_{n\neq 0}\left\{ 2\pi|n|\mbox{Im}\tilde{\tau}
    +\log\left[ (1-e^{-2\pi |n|\Ima\tilde{\tau}})(1-e^{-2\pi |n|\Ima\tilde{\tau}}) \right]\right\}\nonumber\\
  &=-2\log|\eta(\tilde{\tau})|^2
\end{align}
where we use $\zeta(-1)=-1/12$. $\eta(\tau)$ function is defined as
\begin{align}
\eta(\tau)=q^{\frac{1}{24}}\prod_{n=1}^{\infty}(1-q^n)
\end{align}
where $q=e^{2\pi i\tau}$.

After including the first term in $G(s)$ and using $\zeta(0)=-\frac{1}{2}$, we have 
\begin{align}
G^{\prime}(0)=-2\log \left( \sqrt{A \Ima\tilde{\tau}}|\eta(\tilde{\tau})|^2 \right) 
\end{align}
Therefore the free boson partition function is
\begin{align}
Z_{bos}(\tau)=\sqrt{A}\exp\left(\frac{1}{2}G^{\prime}(0)\right)=\frac{1}{\sqrt{\Ima\tilde{\tau}}|\eta(\tilde{\tau})|^2}=\frac{1}{\sqrt{\Ima\tau}|\eta(\tau)|^2}
\end{align}
In the last step, we used the S transformation of the  $\eta(\tau)$ function.

\subsubsection{Twist boundary condition}
If we impose a twisted boundary condition in the $x$ or $y$ direction, the eigenvalues $\omega_{m,n}$ will be modified,
\begin{align}
\omega_{n,m}=\left(\frac{2\pi(n+\lambda_1)}{L_x}\right)^2+\left(\frac{2\pi(m+\lambda_2)}{L_y}\right)^2
\end{align}
where $\lambda_{1,2}$ are the twists along the $x$ and $y$ directions. After considering the twist boundary condition, there is no zero mode anymore and we have 
\begin{align}
\left|\frac{2\pi L_x}{A}\right|^{2s}G(s)=\sum_{m,n\in\mathbb{Z}}\frac{1}{|m+\lambda_2+(n+\lambda_1)\tilde{\tau}|^{2s}}
\end{align}

Using the results in Sec.\ref{epstein}, we have
\begin{align}
\log Z&=-\frac{1}{2}\log\left\{\frac{\thg{\lambda_1-\frac{1}{2}}{\lambda_2-\frac{1}{2}}(\tilde{\tau})}{\eta(\tilde{\tau})}
\frac{\thg{\lambda_1-\frac{1}{2}}{-\lambda_2+\frac{1}{2}}(\tilde{\tau})}{\eta(\tilde{\tau})}\right\}\nonumber\\
&=-\frac{1}{2}\log\left\{\frac{\thg{\lambda_2-\frac{1}{2}}{-\lambda_1+\frac{1}{2}}(\tau)}{\eta(\tau)}
\frac{\thg{-\lambda_2+\frac{1}{2}}{-\lambda_1+\frac{1}{2}}(\tau)}{\eta(\tau)}\right\}
\label{part_torus}
\end{align}
To obtain this result, we perform S transformation $\tau\to-1/\tau$ in the second step.


\subsection{Open cylinder}
\label{cylinder_par}
\subsubsection{Periodic in the $y$ direction}
For the cylinder, if we impose Dirichlet boundary at two boundaries with $\phi(x=0)=\phi(x=L_x)=0$, only half of the modes will remain,
\begin{align}
\left|\frac{2\pi L_x}{A}\right|^{2s}G(s)&=\sum_{m\in\mathbb{Z},n>0}\frac{1}{|m+\frac{n\tilde{\tau}}{2}|^{2s}}\nonumber\\
&=\sum_{n> 0}\left(\sum_{m\in\mathbb{Z}}\frac{1}{|m+\frac{n\tilde{\tau}}{2}|^{2s}}\right)
\end{align}
Using the zeta function regularization technique, we expand $G(s)$ around $s=0$,
\begin{align}
G(s)=\frac{1}{6}s\pi \frac{ \Ima{\tilde\tau} }{2}-s\prod_{n>0}\log\left[ (1-e^{2\pi in\frac{\tilde{\tau}}{2}})^2 \right] +\dotsb
  =-s\log\left( |\eta(\tilde{\tau}/2)|^2 \right) +\dotsb
\end{align}
where the dots denote higher order terms in $s$.

Hence
\begin{align}
Z_{cyl}(\tau)=\exp\left(\frac{1}{2}G^{\prime}(0)\right)=\frac{1}{\eta(-\frac{1}{2\tau})}=\frac{1}{\sqrt{-2i\tau}\eta(2\tau)}
\end{align}
\subsubsection{Twisted boundary condition in the $y$ direction}
In this case, we need to calculate
\begin{align}
\left|\frac{2\pi L_x}{A}\right|^{2s}G(s)&=\sum_{m\in\mathbb{Z},n>0}\frac{1}{|m+\lambda+\frac{n\tilde{\tau}}{2}|^{2s}}\nonumber\\
&=\sum_{n> 0}\left(\sum_{m\in\mathbb{Z}}\frac{1}{|m+\lambda+\frac{n\tilde{\tau}}{2}|^{2s}}\right)
\end{align}
Expanding $G(s)$ around $s=0$, we have 
\begin{align}
  G(s)=\frac{1}{6}s\pi \frac{ \Ima\tilde{\tau}}{2}-s\prod_{n>0}\log\left[ (1-e^{2\pi i\lambda+2\pi in\frac{\tilde{\tau}}{2}}) (1-e^{-2\pi i \lambda+2\pi in\frac{\tilde{\tau}}{2}}) \right]
\end{align}

Therefore the partition function is
\begin{align}
Z&=\tilde{q}^{-1/24}\prod_{n=1}^{\infty}\sqrt{\frac{1}{(1-e^{2\pi i\lambda}\tilde{q}^n)(1-e^{-2\pi i\lambda}\bar{\tilde{q}}^n)}}  \nonumber\\ 
&=e^{\frac{\pi i}{2}(\lambda-\frac{1}{2})}\sqrt{(1-e^{-2\pi i\lambda})}\sqrt{\frac{\eta(-\frac{1}{2\tau})}{\thg{\frac{1}{2}}{\lambda-\frac{1}{2}}(-\frac{1}{2\tau})}}\nonumber\\
&=\sqrt{(1-e^{-2\pi i\lambda})}\sqrt{\frac{\eta(2\tau)}{\thg{\lambda-\frac{1}{2}}{-\frac{1}{2}}(2\tau)}}
\label{part_cyl_s}
\end{align}
where $\tilde{q}=e^{2\pi i\left(\tfrac{1}{-2\tau}\right)}$, $\tau=i\frac{L_x}{L_y}$, $q=e^{2\pi i\tau}$.

For instance, if we impose anti-periodic boundary condition in $y$ direction with $\lambda=1/2$, the partition function becomes
\begin{align}
Z=\sqrt{2}\sqrt{\frac{\eta(-\frac{1}{2\tau})}{\theta_2(-\frac{1}{2\tau})}}=\sqrt{2}\sqrt{\frac{\eta(2\tau)}{\theta_4(2\tau)}}=\sqrt{2}q^{\frac{1}{24}}\prod_{n=1}^{\infty}\frac{1}{(1-q^{2n-1})}
\end{align}

\section{Numerical method for calculating EE in free systems}
\label{numerical_method}
\subsection{Free scalar field theory}
\label{method_n}
The discrete free boson Hamiltonian defined on a lattice reads
\begin{align}
H=\frac{1}{2}\left(\sum_{i=1}^N\Pi_i^2+\sum_{i,j=1}^N\phi_i K_{ij}\phi_j\right)
\end{align}
where $K_{ij}$ is a discretized version of $-\nabla^2 + m^2$. 
The operators satisfy the canonical commutation relations
\begin{align}
[\phi_i,\Pi_j]=i\delta_{ij},\quad [\phi_i,\phi_j]=0,\quad [\Pi_i,\Pi_j]=0
\end{align}
The ground state for this Hamiltonian is 
\begin{align}
|\Psi\rangle  = \mathcal N \sum_{\varphi} e^{-\frac{1}{2}\sum_{ij}\varphi_i K_{ij}^{1/2} \varphi_j}|\varphi\rangle 
\label{gs_free}
\end{align}
where $\varphi=\{\varphi_i\}$ denotes a field configuration, and $\phi_i|\varphi\rangle =\varphi_i |\varphi\rangle$, and $\mathcal N$ is 
a normalization.
We use this wavefunction to calculate the entanglement entropy. For free boson system, there is a very efficient numerical method to calculate the entanglement entropy. Below is a short summary of the method,
while the details can be found in Ref.\thinspace\onlinecite{Casini_rev}. 

The correlation functions for the ground state are
\begin{align}
&\langle \phi_i\phi_j\rangle=\frac{1}{2}K_{ij}^{-1/2}\equiv X_{ij}\nonumber\\
&\langle \pi_i\pi_j\rangle=\frac{1}{2}K_{ij}^{1/2}\equiv P_{ij}
\end{align}
The von Neumann EE for subsystem $A$ can be calculated by using these correlation functions,
\begin{align}
  S_1(A)=\sum_\ell \left(\nu_\ell+\tfrac{1}{2}\right)\log\left(\nu_\ell+\tfrac{1}{2}\right) 
  - \left( \nu_\ell-\tfrac{1}{2} \right)\log\left( \nu_\ell -\tfrac{1}{2} \right)
  \label{s_entr} 
\end{align}
where the $\nu_\ell$ are eigenvalues of $C=\sqrt{X_A P_A}$. $X_A$ and $P_A$ are the correlation functions defined on subregion A.
Similarly, the R\'enyi EE for $n>0$ reads
\begin{align}
S_n=\sum_\ell \frac{1}{n-1}\left[\log((\nu_\ell+1/2)^n-(\nu_\ell-1/2)^n)\right]\,.
\end{align}

\subsubsection{Relativistic boson in $2+1$d}
The Hamiltonian for the relativistic boson in $2+1$d is
\begin{align}
H=\frac{1}{2}\int d^2x \left[\Pi^2+(\nabla\phi)^2+m^2\phi^2\right]
\end{align} 
The corresponding discrete lattice Hamiltonian (on the torus) is
\begin{align}
H=\frac{1}{2}\sum_{i,j}\left[\Pi_{i,j}^2+(\phi_{i+1,j}-\phi_{i,j})^2+(\phi_{i,j+1}-\phi_{i,j})^2+m^2\phi_{i,j}^2\right]
\label{H_discrete}
\end{align}
In the momentum space, the Hamiltonian becomes
\begin{align}
H=\frac{1}{2}\int_{-\pi}^{\pi} dk_1\int_{-\pi}^{\pi} dk_2 \Pi(k)\Pi(-k)+\left[4-2\cos(k_1)-2\cos(k_2)+m^2\right]\phi(k)\phi(-k)
\end{align}
The two point correlation functions are
\begin{align}
&\langle\phi_{i,j}\phi_{i+n_1,j+n_2}\rangle=\frac{1}{8\pi^2}\int_{-\pi}^{\pi} dk_1\int_{-\pi}^{\pi}dk_2\frac{\cos(k_1n_1)\cos(k_2n_2)}{\sqrt{4-2\cos(k_1)-2\cos(k_2)+m^2}}\nonumber\\
&\langle\pi_{i,j}\pi_{i+n_1,j+n_2}\rangle=\frac{1}{8\pi^2}\int_{-\pi}^{\pi} dk_1\int_{-\pi}^{\pi}dk_2{\cos(k_1n_1)\cos(k_2n_2)}{\sqrt{4-2\cos(k_1)-2\cos(k_2)+m^2}}
\end{align}
we can use these correlation functions to construct $C$ matrix and calculate EE.

\subsection{Numerical method for the free Dirac fermion in $2+1$ dimensions}

Here we follow the method in Ref.\ \onlinecite{Peschel2003} and show a simple example for massless Dirac fermion. The Hamiltonian in the momentum space is
\begin{align}
H_D=\int_{BZ}\frac{d^2k}{(2\pi)^2}\Psi^{\dag}({\bf k})\mathcal{H}_D({\bf k})\Psi({\bf k})
\end{align}
where $\Psi^{\dag}({\bf k})=(c_{\bf k}^{\dag}, d_{\bf k}^{\dag})$ is a two component spinor, BZ stands for the first Brillouin zone, $-\pi<k_x\leq \pi$ and $-\pi<k_y\leq \pi$. The one-particle lattice Dirac Hamiltonian $\mathcal{H}_D({\bf k})$ takes this form
\begin{align}
\mathcal{H}_D({\bf k})=h_1\sigma_x + h_3\sigma_z
\end{align}
with $h_1=\cos(k_x)$ and $h_3=\cos(k_y)$. The Dirac points are at $(\pm \pi/2,\pm \pi/2)$ and there are four Dirac cones in the first Brillouin zone.

$\mathcal{H}_D({\bf k})$ can be diagonalized by a unitary transformation $V^{-1}\mathcal{H}_DV=M$ where $M$ is the diagonal matrix with eigenvalues $E({\bf k})_{\pm}=\pm\sqrt{\cos(k_x)^2+\cos(k_y)^2}$. The V matrix equals to  
\begin{align}
V=&\frac{1}{\sqrt{(2\cos(k_y)^2+2\cos(k_x)^2-2\cos(k_y)\sqrt{\cos(k_y)^2+\cos(k_x).^2}}}\nonumber\\
&\times\begin{pmatrix}
\cos(k_x) &-\cos(k_y)+\sqrt{\cos(k_y)^2+\cos(k_x)^2}  \\
-\cos(k_y)+\sqrt{\cos(k_y)^2+\cos(k_x)^2} &-\cos(k_x)
\end{pmatrix}
\end{align} For the ground state with lower band fully filled, the correlation functions in momentum space equal to
\begin{align}
\langle c_{\bf k}^{\dag}c_{\bf k}\rangle=V_{12}^2,\quad \langle d_{\bf k}^{\dag}d_{\bf k}\rangle=V_{22}^2,\quad \langle c_{\bf k}^{\dag}d_{\bf k}\rangle=V_{12}V_{22}
\end{align}
Using the above equations, we can construct the correlation function matrix $C_E$ for the subsystem A and therefore the von Neumann EE is
\begin{align}
S_{vN}=-\sum_{\ell}\left[ \nu_\ell \log \nu_\ell+(1-\nu_\ell)\log(1-\nu_\ell)\right]
\end{align}
where $\nu_\ell$ is the eigenvalue for $C_E$. Similarly, the R\'enyi entropy is
\begin{align}
S_n=\frac{1}{1-n}\sum_{\ell} \log\left[(1-\nu_\ell)^n+\nu_\ell^n\right]
\end{align}

\bibliographystyle{apsrev}  
\bibliography{EE_twist}
\end{document}